\documentclass[fleqn,usenatbib]{mnras}

\usepackage{newtxtext,newtxmath}
\usepackage{xcolor}
\usepackage{multirow}
\usepackage{natbib}

\usepackage[T1]{fontenc}

\DeclareRobustCommand{\VAN}[3]{#2}
\let\VANthebibliography\thebibliography
\def\thebibliography{\DeclareRobustCommand{\VAN}[3]{##3}\VANthebibliography}

\usepackage{graphicx}	
\usepackage{amsmath}	
\usepackage{threeparttable}
\usepackage{caption}
\usepackage{subcaption}
\usepackage{array}
\usepackage{hyperref}
\usepackage{tablefootnote}
\newcommand{\seff}{$S_{\textrm{eff}}$}
\newcommand{\teff}{$T_{\textrm{eff}}$}
\newcommand{\rp}{$R_{\textrm{p}}$}
\newcommand{\teq}{$T_{\textrm{eq}}$}
\newcommand{\logg}{$\rm{\log g}$}

\title[Habitability and structural composition]{Exploring the habitability and interior composition of exoplanets lying within the extended habitable zone}

\author[Deb et al.]{
Sushmita Deb$^{1}$\thanks{E-mail: mail.sushmitadeb@gmail.com},
Kaushal Sharma$^{2,3}$,
Samrat Biswas$^{1}$, and
Biman Jyoti Medhi$^{1}$
\\
$^{1}$Department of Physics, Gauhati University, Jhalukbari, Guwahati 781014, India\\
$^{2}$Forensic Science Laboratory Uttar Pradesh, Ghaziabad 201204, India\\
$^{3}$Inter University Centre for Astronomy and Astrophysics, Pune 411007, India\\
}

\pubyear{\the\year{}}

\begin{document}

\label{firstpage}
\pagerange{\pageref{firstpage}--\pageref{lastpage}}
\maketitle

\begin{abstract}
Studying the habitability, internal structure and composition of exoplanets is crucial for understanding their potential to sustain life beyond our solar system. Characterizing planetary structures and atmospheric evolution provides valuable insights into surface conditions and the long-term habitability of these planets. In this study, we present a comprehensive analysis of exoplanets spanning from super-Earths to mini-Neptunes ($R_{\textrm{p}}$ $\leq$ 4 $R_{\oplus}$ and $M_{\textrm{p}}$ $\leq$ 15 $M_{\oplus}$) located within the extended habitable zone, along with parameterization of their host stars. We find that the planets in our sample orbit M dwarf stars and are tidally locked to them. Using archival photometric data from Gaia, Pan-STARRS1, 2MASS, and WISE, we estimate the atmospheric and physical parameters of the host stars. We also model the interior structure of these planets to infer their possible compositions. Additionally, under the assumption that these exoplanets can accrete a gaseous layer, we model the envelope fraction of the habitable exoplanets. With an Earth-like rocky composition,  LHS 1140 b and TOI-1452 b can hold onto negligible amount of their initial gas layer. However, sustaining a sufficient amount of atmosphere over time, the planets LP 791-18 c, LTT 3780 c and K2-18 b are likely to be water worlds. The models suggest a water rich composition for TOI-1266 c without any significant amount of atmosphere. Modeling interior compositions and atmospheric escape scenarios allow us to assess the potential habitability of these planets by evaluating the likelihood of surface liquid water and the retention of stable atmospheres.

\end{abstract}




\section{Introduction}

With significant advancements in the detection and characterisation of exoplanets, the prospect of their potential habitability has also gained interest in recent times. The pioneering works by \citet{hart1978evolution, kasting1988climate, kasting1988runaway, whitmire1991habitable} have defined habitability of extrasolar planets based on the availability of liquid water on their surface, as it is considered essential for life as we know it. The ability of a planet to retain surface water is influenced by various factors, such as the star’s surface temperature, its evolution, the distance between the star and the planet, etc. These conditions define a specific region around the star known as the habitable zone (HZ) \citep{huang1959occurrence, huang1960life}, where planets are most likely to be potentially habitable. The inner edge of this region, known as the Inner Habitable Zone (IHZ), is influenced by the processes like the runaway greenhouse or moist greenhouse effect \citep{kasting1993habitable, kopparapu2013habitable}. If a planet lies too close to its host star, intense infrared (IR) radiation can trigger a runaway greenhouse effect, leading to excessive atmospheric heating that results in the complete vaporization of surface water. On the other hand, the outer edge of the HZ (OHZ) is defined as the distance beyond which surface water would freeze due to insufficient heat from the host star. 

However, specific conditions, such as Earth-like cold trapping of water vapor \citep{wandel2018biohabitability} or significant cloud coverage, can shift the IHZ closer to the star, potentially up to 0.5AU for a Sun-like star \citep{kasting1993habitable, yang2016differences}. Tidal heating is another factor that can alter the boundaries of the classical HZ. Many exoplanets in close orbits around M-dwarf stars are tidally locked to their host stars and experience extreme volcanic activity due to continuous tidal heating \citep{driscoll2015tidal}. Under such intense conditions, the day side of these planets experiences significant degassing, preventing the retention of surface liquid water. However, the night side can maintain frozen water if the heat transfer from the day side is insufficient to melt and evaporate the ice \citep{wandel2023habitability}. If enough heat is transported, liquid water may be present in certain regions of these planets \citep{wandel2018biohabitability}. These water bodies, shielded from stellar radiation on the night side, can shift the IHZ closer to the host star \citep{wandel2023extended}. 

Near the OHZ, volcanic outgassing can lead to the formation of thick $\textrm{CO}_2$ atmospheres on planets, which serve as greenhouse agents that warm the planet and push the OHZ boundary outward \citep{kopparapu2013habitable, forget2013probability}. This $\textrm{CO}_2$-induced warming effect broadens the range within which liquid water can exist, enhancing the habitability potential for planets near the OHZ. \citet{ojha2022liquid} highlight that geothermal heat from radiogenic elements can cause ice sheets to melt and sustain surface liquid water at temperatures above freezing point, even for planets situated near or beyond the OHZ. This process may also expand the habitable zone to include regions outside the classical OHZ.

The habitability of exoplanets is also governed by their interior structure, composition, and long-term atmospheric evolution \citep{noack2013interior, noack2014can, cockell2016habitability, dorn2017assessing, van2019exoplanet}. Previous studies, aimed to investigate the internal structure of exoplanets \citep{valencia2006internal, valencia2007radius, sotin2007mass, fortney2010interior, becker2013dynamical, howe2014mass, kellermann2018interior, wang2019enhanced, baumeister2020machine, Daspute_2025}, assume a typical four layers model: an iron-rich core, a silicate mantle, an ice and ocean layer, and a gas-rich atmospheric envelope primarily consisting of hydrogen and helium. Various combinations of planetary mass and radius, along with factors such as planetary dynamics, can produce a wide range of layer proportions leading to a diversified population of exoplanets with differing internal structures. The presence and amount of volatiles, such as water and gases, significantly influence the climate conditions on these planets. Processes like volcanic outgassing or the cooling of magma oceans \citep{noack2017volcanism, oosterloo2021role} can have a notable impact on atmospheric dynamics, especially for smaller planets, which in turn affects their potential habitability. Studying the internal structure and composition of exoplanets is key to understanding how chemical elements move and cycle between the different layers of a planet \citep{dorn2017assessing}. This insight is essential for determining a planet's ability to support a layer of liquid water enriched with minerals necessary for sustaining life.

The ability of exoplanets to retain their primordial atmospheres is another key factor in determining their long-term habitability. Observations from exoplanet surveys and subsequent studies have shown strong evidence of a bimodal distribution in the radii of short-period planets, revealing two distinct peaks at approximately 1.3$R_{\oplus}$ and 2.4$R_{\oplus}$ \citep{fulton2017california}. This distribution suggests the presence of two separate populations: rocky super-Earths and less dense, gaseous sub-Neptunes, separated by a radius gap or valley around $1.5\,-\,2\,R_{\oplus}$. One mechanism proposed to explain this radius gap is photoevaporation, where close-in planets experience atmospheric escape and mass loss due to extreme ultraviolet (EUV) and soft X-ray radiation from their host stars \citep{2004Icar..170..167Y, murray2009atmospheric, owen2013kepler}. Alternatively, the core-powered mass-loss mechanism \citep{ginzburg2018core, gupta2019sculpting} suggests that energy from a planet’s cooling core drives atmospheric loss \citep{ginzburg2018core, gupta2019sculpting}. In a gas-rich environment, this mechanism can also lead to the accretion of H/He atmospheres, which act as a `thermal blanket', trapping the core’s thermal energy. Considering these mass transfer processes, exoplanets that lose their atmospheres entirely evolve into super-Earths, while those that retain part of their atmospheres are classified as sub-Neptunes. The occurrence rate of exoplanets within this valley is notably lower, and the gas-depleted formation model \citep{lee2021primordial}, in combination with the aforementioned mass and heat transfer mechanisms, is thought to contribute to the formation of this radius gap. 

In this paper, we focus on a subset of known exoplanets with $R_{\textrm{p}}\,\leq\,4R_{\oplus}$ and $M_{\textrm{p}}\,\leq\,15 M_{\oplus}$ that are located within the extended habitable zone. We parameterize their host stars and investigate the structural and atmospheric properties of the selected exoplanets. Section~\ref{2} details the sample selection process and the data analysis strategies employed to determine stellar parameters, study interior structures, and evaluate atmospheric conditions of the sample exoplanets. Section~\ref{3} presents the findings from our structural analysis and discusses the atmospheric evolution of the exoplanets. Section~\ref{4} describes the atmospheric characterization prospect for the sample exoplanets through Transmission Spectroscopic Metric (TSM) which could be useful in prioritizing and scheduling future observing sessions dedicated to characterize the exoplanetary atmosphere. Finally, Section~\ref{5} summarizes our key findings and provides the conclusions.

\section{Sample selection and data analysis} \label{2}

To study exoplanet habitability, we download the list of exoplanets from NASA Exoplanet Archive\footnote{\href{https://exoplanetarchive.ipac.caltech.edu/ on 18 November, 2024}{NASA Exoplanet Archive}}, applying sample selection criteria with a radius $\leq 4R_{\oplus}$ and a mass $\leq\,15M_{\oplus}$. We opt for these thresholds to focus on the habitability of terrestrial core planets with comparable Earth radius. The selection criteria provided with 881 planets with number of repetitive entries from multiple studies and observations. We identify a total of 339 exoplanets after removing duplicates and filtering out planets with missing measurements on radius, semi-major axis and stellar effective temperature.

\begin{table*}
    \centering
    \caption{The planetary parameters and insolation flux (\seff) for all the habitable zone exoplanets. The parameters listed are: $P$ (orbital period), $a$ (semi-major axis), $R$ (radius), $M$ (mass), $\rho$ (density), and $T_{\textrm{eq}}$ (equilibrium temperature). The sources for the planetary parameters are provided in the last column.}
    \label{tab1}
    \begin{tabular}{llcccccccl}
    \hline
    Planet   & Host star    & $P$         & $a (a_{\odot}, 10^{-3})$          & $R$     & $M$     & $\rho$ & $T_{\textrm{eq}}$    & \seff{} & Source\\ 
    & & (days) & (au) & (R$_{\oplus}$) & (M$_{\oplus}$) & ($\rho_{\oplus}$) & (K) & (S$_{\oplus}$) & \\
    \hline
    TRAPPIST-1 h & TRAPPIST-1 & 18.76$\pm$0.00008 & 62 $\pm$0.5 & $0.773^{+0.026}_{-0.027}$ & $0.331^{+0.056}_{-0.049}$ & $0.719^{+0.117}_{-0.102}$   & 169.2$\pm$2.4  & 0.15$\pm$0.01 &  1,2,3 \\
    TRAPPIST-1 g & TRAPPIST-1 & 12.35$\pm$0.00005 & 46$\pm$0.4 & $1.148^{+0.032}_{-0.033}$ & $1.148^{+0.098}_{-0.095}$ & $0.759^{+0.034}_{-0.033}$   & 194.5$\pm$2.7  & 0.26$\pm$0.01 & 1,2,3\\
    TRAPPIST-1 f & TRAPPIST-1 & 9.20$\pm$0.00003   & 38$\pm$0.3 & $1.046^{+0.029}_{-0.030}$ & $0.934^{+0.080}_{-0.078}$ & $0.816^{+0.038}_{-0.036}$ & 214.5$\pm$3.0  & 0.38$\pm$0.02 & 1,2,3\\
    TRAPPIST-1 e & TRAPPIST-1 & 6.10$\pm$0.00001  & 29$\pm$ 0.7 & $0.910^{+0.026}_{-0.027}$ & $0.772^{+0.079}_{-0.075}$ & $1.024^{+0.076}_{-0.070}$   & 246.1$\pm$3.5  & 0.65$\pm$0.05 & 1,2\\
    TRAPPIST-1 d & TRAPPIST-1 & 4.05$\pm$0.00003  & 22$\pm$0.2 & 0.784$\pm$0.023 & $0.297^{+0.039}_{-0.035}$ & $0.616^{+0.067}_{-0.062}$  & 282.1$\pm$4.0  & 1.13$\pm$0.05 & 1,2,3\\
    TRAPPIST-1 c & TRAPPIST-1 &  2.42$\pm$0.00002  & 15$\pm$0.1 & $1.095^{+0.030}_{-0.031}$  & $1.156^{+0.142}_{-0.131}$ & $0.883^{+0.083}_{-0.078}$  & 334.8$\pm$4.7  & 2.25$\pm$0.09 & 1,2,3\\
    TOI-1452 b & TOI-1452  & 11.06$\pm$0.00002  & 61$\pm$3.0      & 1.672$\pm$0.071 & 4.820$\pm$1.300  & 1.010$\pm$0.320     & 326.0$\pm$7.0   & 1.95$\pm$0.19 & 4\\
    TOI-1266 c & TOI-1266  & 18.80$\pm$0.00005 & 104$\pm$1.6     & 2.130$\pm$0.120  & 2.880$\pm$0.800  & 0.290$\pm$0.100     & 354.0$\pm$16.0    & 2.72$\pm$0.49 & 5\\
    LTT 3780 c & LTT 3780  & 12.25$\pm$0.00007 & 76$\pm$3.4     & 2.420$\pm$0.100  & 6.290$\pm$0.630  & 0.440$\pm$0.079    & 397.0$\pm$39.0    & 2.86$\pm$0.22 & 6\\
    LP 791-18 c & LP 791-18 & 4.98$\pm$0.00005 & 29$\pm$1.0   & 2.438$\pm$0.096 & 7.100$\pm$0.700  &  0.117$\pm$0.014     & 370.0$\pm$30.0    & 2.59$\pm$0.34 & 7,8\\
    LHS 1140 b & LHS 1140   & 24.73$\pm$0.00002  & 94$\pm$1.7     & 1.730$\pm$0.025  & 5.600$\pm$0.190  & 1.070$\pm$0.034     & 226.0$\pm$4.0    & 0.44$\pm$0.03 & 9\\
    K2-18 b\  & K2-18    & 32.94$\pm$0.00001 & 159$\pm$0.5     & 2.610$\pm$0.087  & 8.630$\pm$1.350  & 0.480$\pm$0.094    & 284.0$\pm$15.0 & 0.98$\pm$0.06 & 10,11\\
    \hline
    \end{tabular}%
\begin{tablefootnote}
   1   1: \citet{Agol_2021}; 2: \citet{delrez2018early}; 3: \citet{grimm2018nature}; 4: \cite{cadieux2022toi}; 5: \citet{cloutier2024masses}; 6: \citet{nowak2020carmenes}; 7: \citet{crossfield2019super}; 8: \citet{peterson2023temperate}; 9: \citet{cadieux2024new}; 10: \citet{2018AJ....155..257S}; 11: \citet{benneke2019water}
\end{tablefootnote}
\end{table*}

\subsection{Identifying Exoplanets within the Habitable Zone}

The classical HZ boundaries, as defined by \citet{kasting1993habitable} and later by \citet{kopparapu2013habitable}, delineate a specific region around stars with different temperatures where conditions may allow planets to sustain liquid water on their surface. The inner and outer edges of the HZ mark the closest and farthest distances to the star, respectively, beyond which the planets are incapable of sustaining surface liquid water due to insufficient stellar radiations. 

Many close-in planets around M-dwarf stars get tidally locked, facing extreme temperatures at its two opposite phases. Using the tidal locking relations from \citet{burns1986satellites} and \citet{gladman1996synchronous} and the main-sequence relations between luminosity, mass, and surface temperature, \citet{wandel2023extended} related the insolation flux received by tidally locked planets with the effective temperature of the host star as $ \frac{S_{\textrm{r}}}{S_{\textrm{E}}} \sim \, 7.3 (\frac{T_{*}}{T_{\odot}})^{5.7}$, where $S_{\textrm{r}}$ is the insolation flux received by the planet and $T_{*}$ is the effective temperature of the host star. Planets with eccentric orbits, lying within this radius, face intense volcanic eruption and extreme conditions on the day-side, which leads to significant degassing and elevated surface temperatures, as highlighted by \citet{driscoll2015tidal}. The absence of magnetic field on these planets also contributes to the gradual erosion of their atmospheres by stellar winds \citep{barnes2017tidal}. However, \citet{wandel2023extended,wandel2023habitability} has also shown that if the heat transport from the day side is low enough, the night side of these planets can maintain frozen ice sheets and cooler surface temperatures. Taking the moist greenhouse inner edge limit of \citet{kopparapu2013habitable} as the radiative energy sufficient to melt the night-side ice sheets, the author has expanded the classical HZ boundary, marking the IHZ for several values of heat transported from the day side to the night side of tidally locked exoplanets

\citet{wandel2023extended} also suggested a lower flux value for the outer boundary of the EHZ, based on the evidence of an intra-glacial lake in the south polar region of Mars \citep{2018Sci...361..490O}. They estimated the annual flux received by the polar region of Mars as $<(\cos(i)>\,\sim\,0.3$ of the irradiation at the orbit of Mars, and arrived at a value of 0.1$S_{\textrm{E}}$, which marks the outer boundary of the EHZ.
 
\begin{figure*}
    \centering
    \includegraphics[width=0.9\linewidth]{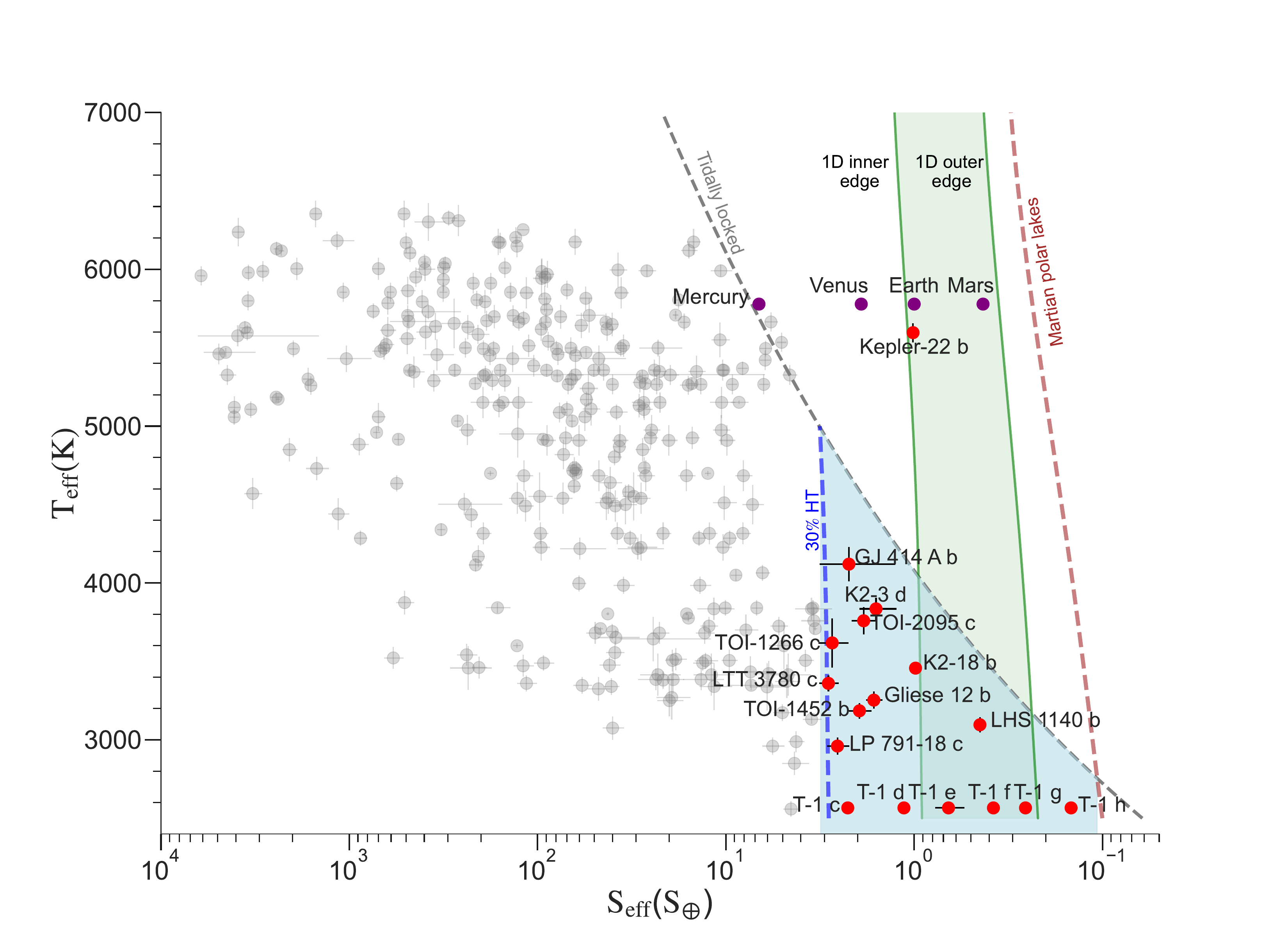}
    \caption{Distribution of 339 planets in the stellar effective flux (\seff{}) versus stellar temperature (\teff{}) space. Red solid circles indicate the 17 planets orbiting within the EHZ, while gray points depict the remaining planets. Solar system planets are marked with purple circles. The green solid lines correspond to the inner and outer edges of the classical 1D habitable zone, as defined by \citet{kopparapu2013habitable}. The blue dashed line represents the inner edge of the Extended Habitable Zone (EHZ) around M dwarf stars, assuming $30\%$ heat transport efficiency \citep{wandel2023habitability}. The brown curve marks the outer edge of the EHZ, determined by the melting of ice sheets driven by geothermal heat \citep{wandel2023extended}. The gray dashed line indicates the tidally locked radius around M dwarf stars. The green shaded region defines the classical HZ and the blue shaded region highlights the extended HZ for tidally locked exoplanets}.
    \label{fig:1}
\end{figure*}

Fig.~\ref{fig:1} illustrates the various regions of the Extended Habitable Zone (EHZ) on a stellar effective temperature versus insolation plot. The green solid lines represent the classical 1D inner and outer habitable zone (IHZ and OHZ) boundaries as defined by \citet{kopparapu2013habitable}. The blue dashed lines indicate the extended IHZ boundary around M dwarf stars, corresponding to $30\%$ heat transport from the day side to the night side of tidally locked exoplanets. The extended outer habitable zone (OHZ), also referred to as the Martian polar lakes region \citep{wandel2023extended}, is marked by the brown dashed line. The gray curve represents the tidally locked radius as defined by \citet{wandel2023extended}.

For the total population of 339 exoplanets, we first compute the luminosity of the star, $L_*$, using equation 16.1 from \citet{cockell2020astrobiology}: 
\begin{eqnarray}\label{eq:luminosity}
L_{*}=4\pi R^{2}_{*}T^{4}_{*},
\end{eqnarray}
where $R_{*}$, and $T_{*}$ are the radius and temperature of the host star respectively. $R_{*}$ and $T_{*}$ values are adopted from NASA Exoplanet archive. We further determine the insolation flux, $S_{\textrm{eff}}$, using equation 16.3 from \citet{cockell2020astrobiology}
\begin{eqnarray}\label{eq:insolation_flux}
S_{\textrm{eff}}=\frac{L_{*}}{4\pi a^2},
\end{eqnarray}
where $a$ is the semi-major axis for the exoplanet orbit. Fig.~\ref{fig:1} shows that only 17 planets (represented by the red solid circles) lie within the EHZ, out of the population of 339 exoplanets (rest of the planets are denoted by gray circles). Of these planets GJ 414 A b has 32$\%$ of mass measurement error and 46$\%$ of radius measurement error, while Kepler-22 b, K2-3 b, TOI-2095 c and Gliese 12 b have no known mass measurement errors. Therefore, we omit these five planets and are left with 12 exoplanets for our further analysis.  

All the remaining 12 exoplanets orbit M-dwarf stars with effective temperatures ranging from 2900\,K to 3900\,K and are likely tidally locked to their host stars. Exoplanets such as TOI-1266 c, LTT 3780 c, LP 791-18 c, and TOI-1452 b lie well within the extended habitable zone (HZ), suggesting that their rotation is synchronously locked with their parent stars. The planetary parameters and calculated insolation flux for these planets are presented in Table~\ref{tab1}. The observed bulk density of LP 791-18 c was not reported in the literature; therefore we calculated it using $\rho=\frac{3M}{4\pi R^{3}}$\citep{2024A&A...686A.296M}.

In the following sections, we examine the fundamental stellar parameters, structural composition, and atmospheric evolution of all the exoplanets listed in Table~\ref{tab1}, except the for the Trappist-1 system. The Trappist-1 planetary system has been extensively studied by \citet{Agol_2021}, showing no evidence of volatiles for Trappist-1 b, c and d, and small amounts of water content for the rest of Trappist-1 e, f, g and h planets. Therefore, we focus our analysis on the remaining six exoplanet systems.

\subsection{Stellar Parameters from Archival Photometry}
We estimate the spectral types (SpT) of the host stars by comparing their Gaia DR3 photometric magnitudes and parallaxes \citep{2023A&A...674A...1G} with the color and absolute magnitude relations provided by \citet{2019AJ....157..231K}. Using six distinct color-absolute magnitude relations from \citet{2019AJ....157..231K}, we verify the consistency of the derived properties and classify the host stars as M dwarfs, with their specific subclasses listed in the second column of Table~\ref{tab2}. Both \citet{1995AJ....110.1838R} and \citet{2017Natur.544..333D} classified LHS 1140 as an M4.5V spectral type based on low- and high-resolution spectroscopy. Our estimated spectral type for the star aligns well with these literature values. For TOI-1452, \citet{cadieux2022toi} estimated a spectral type between M4 and M4.5 based on high-resolution spectroscopy and empirical temperature-color relations. Also, they independently determined a spectral type of M3.7$\pm$0.6 by comparing Gaia DR2 colors with the spectral type relations of \citet{2019AJ....157..231K}. Our estimated spectral type is consistent with the average of the values determined by \citet{cadieux2022toi}. For the host star TOI-1266, \citet{2020A&A...642A..49D} and \citet{2020AJ....160..259S} independently determined spectral types of M3 and M2, respectively, based on spectroscopic analysis and cross-matching with archival spectra. Our assigned spectral type is consistent with these determinations. For LP 791-18, our estimated spectral type shows excellent agreement with \citet{crossfield2019super}, who reported a type of M(6.1$\pm$0.7)V using the color-SpT relation from \citet{2019AJ....157..231K}. Similarly, for LTT 3780, \citet{2003AJ....126.3007R} and \citet{2005A&A...442..211S} determined spectral types of M3.5V and M4, respectively, based on spectroscopic observations. Our estimate aligns well with the literature values. Finally, for K2-18, the spectral type has been reported as M3V by \citet{2019AJ....158...87D} and M2.5 by \citet{2019A&A...625A..68S}, both based on spectroscopic observations. Our estimated spectral type is in agreement with the cited literature values.

\begin{table*}
\centering
\caption{Derived stellar parameters and associated errors for the host stars in our sample. }
\label{tab2}
     \begin{tabular}{llccrcc}
        \hline
        Star & Spectral Type       & \teff{} (K)      & $\log g$ (dex)  & [Fe/H] (dex)  & $L_{*}(L_{\odot}, 10^{-3}$)   & Age (Gyr)  \\
        \hline
        LHS 1140   & M(4.90$\pm$0.50) & 3100$\pm$50 & 5.50$\pm$0.25 & $-1.00\pm0.25$  & 4.556$\pm$0.147 & $6.63^{10.9}_{2.1}$ \\
        TOI-1452   & M(4.60$\pm$0.47) & 3100$\pm$50 & 5.00$\pm$0.25 & $-0.50\pm0.25$  & 7.613$\pm$0.011 & $5.75^{+10.8}_{-1.7}$ \\
        TOI-1266   & M(2.00$\pm$0.37) & 3600$\pm$50 & 6.00$\pm$0.25 & $-1.00\pm0.25$  & 28.79$\pm$0.068 & $6.16^{+10.6}_{-1.8}$ \\
        LP 791-18  & M(6.16$\pm$0.37) & 2900$\pm$50 & 5.50$\pm$0.25 & $0.30\pm0.12$   & 2.227$\pm$0.001 & $1.99^{+3.5}_{-0.6}$\\
        LTT 3780   & M(3.50$\pm$0.41) & 3300$\pm$50 & 6.00$\pm$0.25 & $0.00\pm0.20$   & 16.50$\pm$0.079 & $6.45^{+10.9}_{-2.1}$   \\
        K2-18      & M(2.50$\pm$0.37) & 3400$\pm$50 & 5.50$\pm$0.25 & $0.00\pm0.20$   & 29.91$\pm$0.016 & $6.60^{+11.1}_{-2.3}$\\
        \hline
    \end{tabular}
\end{table*}

The effective temperature (\teff), surface gravity (\logg), metallicity, and luminosity of the stars are determined through the photometric Spectral Energy Distribution (SED) of each individual star. We utilize the Virtual Observatory SED Analyser\footnote{\href{http://svo2.cab.inta-csic.es/theory/vosa/}{VOSA}} (VOSA) \citep{2008A&A...492..277B} to model the SED fit of the host stars. We used photometry fluxes spanning a broad wavelength range, from optical data from Gaia DR3 \citep{2016A&A...595A...1G,2023A&A...674A...1G} and Pan-STARRS1 DR2 \citep{2016arXiv161205560C}, to infrared data from the Two Micron All Sky Survey (2MASS) \citep{2006AJ....131.1163S} and the Wide-field Infrared Survey Explorer (WISE) \citep{2010AJ....140.1868W}. We provide the VOSA input interface with spatial coordinates, distance (in parsecs) from Gaia DR3 parallax, and individual extinction values for the exoplanets, all obtained from the SINGLE OBJECT search facility of the Gaia Archive\footnote{\href{https://gea.esac.esa.int/archive/}{Gaia Archive}}. VOSA applies extinction corrections to the observed fluxes and provides the adjusted values. It then selects an appropriate theoretical stellar spectral model and generates the corresponding synthetic photometry. Finally, a reduced $\chi^{2}$ test is performed between the synthetic and corrected flux values to determine the best-fit SED parameters. We use the BT-NextGen(AGSS2009) \citep{2009ARA&A..47..481A} and BT-NextGen(GNS93) \citep{1993A&A...271..587G} theoretical models to generate the synthetic photometry. 

The SED fit returns the best-fitting values for the effective temperature (\teff), surface gravity (\logg), metallicity ([Fe/H]), and luminosity ($L_{*}$) of the host stars. These estimated stellar parameters are used as Gaussian priors to derive the approximate age of the individual star using MIST evolutionary tracks \citep{2016ApJS..222....8D, 2016ApJ...823..102C}. The best-fitting stellar parameters for all six stars are provided in Table~\ref{tab2}. Below, we discuss the estimated stellar parameters for each host star and compare them with the values available in the literature.

\textbf{LHS 1140}: Our estimated \teff{} is in good agreement with the average temperature values of 3096$\pm$48\,K and 3150\,K, as determined by \citet{cadieux2024new} and \citet{2021MNRAS.504.5788R}, respectively, based on their spectroscopic analyses. Our estimated metallicity and \logg{} values are also consistent with their values of $-0.15$\,dex and $-0.13$dex for metallicity, and 5.04\,dex and 5.01\,dex for \logg. Our estimated luminosity value is in good agreement with that of \citet{cadieux2024new}. We estimate the approximate age of the star to be around 6.6 Gyr, which is consistent with the value suggested by \citet{2017Natur.544..333D}, who proposed an age greater than 5 Gyr.

\textbf{TOI-1452}: Previous works by \citet{cadieux2022toi} and \citet{2022ApJS..259...35A} have assigned \teff{} values of 3185$\pm$50\,K and 3336\,K, respectively, for this source through spectroscopic observations and analysis. However, our estimate aligns more closely with \citet{cadieux2022toi}, who determined a \teff{} of 3100$\pm$50\,K through SED fitting, which matches our \teff{} estimate. Our estimated \logg{} value also closely agrees with their reported values of 4.95\,dex and 4.69\,dex, respectively. Also, the luminosity value estimated from our SED fitting is consistent with the value reported by \citet{cadieux2022toi}. We could not find an estimated age for this source in the literature.

\textbf{TOI-1266}: Several archival studies have reported the \teff, \logg, and [Fe/H] values for this source, employing empirical magnitude-temperature relations as well as photometric and spectroscopic analyses. \citet{2020A&A...642A..49D} presented temperature estimates of 3548$\pm70$\,K and 3570$\pm$100\,K, derived from spectra obtained using the HIRES \citep{1994SPIE.2198..362V} and Tillinghast Reflector Echelle Spectrograph (TRES) \citep{Furesz:2008} instruments, respectively. They also reported \teff{} values of 3600$\pm$150\,K and 3533$\pm$45\,K from independent SED analyses. For the [Fe/H] values, three estimates were provided: $-0.24\pm0.09$\,dex, $-0.03\pm0.18$\,dex, and $-0.5\pm0.5$\,dex, based on HIRES spectra, TRES spectra, and photometric SED fitting, respectively. \citet{2020AJ....160..259S} also reported \teff{} values of 3563$\pm$77\,K and 3573$\pm$37\,K, derived from spectroscopic analysis and model-SED fitting, respectively. Their corresponding metallicity estimates were $-0.121\pm0.13$\,dex and $-0.08\pm0.11$\,dex. \citet{2021A&A...656A.162M} assigned a \teff{} of 3686\,K and a metallicity of $-0.43$\,dex based on their spectroscopic analysis. Our estimated \teff{} and [Fe/H] values are in good agreement with the values reported by these studies. However, our estimated \logg{} deviates slightly from the range of 4.7\,dex to 4.8\,dex reported in the literature. This discrepancy may be attributed to differences in the model spectra used in our SED analysis, as well as potential mismatches between the adopted model spectra and the photometric flux values.

\textbf{LP 791-18}: Our estimates of \teff{} and \logg{} for LP 791-18 are in good agreement with the reported values of 2960$\pm$55\,K and 5.115$\pm$0.094\,dex, respectively, as given by \citet{crossfield2019super}, who derived these values from photometric relations. The authors also utilized various calibrated photometric relations and reported an average [Fe/H] value of $-0.09\pm0.19$\,dex, which is consistent with our estimated [Fe/H] value. Their luminosity estimate of $0.00201\pm0.00045$ aligns well with our result. Based on the inferred mass, radius, luminosity, and the galactic space velocity of LP 791-18, \citet{crossfield2019super} suggested that the star is likely a few Gyr old. Our estimate of the star's age is 1.99 Gyr.

\textbf{LTT 3780}: \citet{nowak2020carmenes}, \citet{2021A&A...656A.162M}, \citet{2021MNRAS.504.5788R}, and \citet{2024A&A...682A..66B} reported \teff{} values of $3360\pm51$\,K, 3433\,K, 3354\,K, and 3358\,K, respectively. \citet{nowak2020carmenes} employed model-SED fitting, while \citet{2021A&A...656A.162M} and \citet{2021MNRAS.504.5788R} applied spectral synthesis techniques and photometric calibrations, and \citet{2024A&A...682A..66B} used spectroscopic analysis to determine their parameters. Our estimated temperature is in agreement with these reported values. Regarding metallicity, the above studies reported a range of [Fe/H] values of $0.09$\,dex, $-0.27$\,dex, $0.13$\,dex, and $0.06$\,dex, respectively, and our estimated metallicity is closer to the values reported by \citet{nowak2020carmenes} and \citet{2024A&A...682A..66B}. However, our estimated \logg{} from SED fitting is slightly higher and does not fully match the reported \logg{} values of 4.81\,dex, 4.73\,dex, 4.83\,dex, and 4.85\,dex. This discrepancy may stem from a small mismatch between the model spectra and the flux values in the SED analysis. The luminosity value we estimate is in good agreement with that reported by \citet{nowak2020carmenes}. 

\textbf{K2-18}: \citet{2018A&A...620A.180R} reported \teff{} and \logg{} values of 3500\,K and 5.0\,dex, respectively, for K2-18, based on a comparison of high-resolution spectra with theoretical synthetic models. \citet{2018AJ....155...84W} assigned \teff{}\,=\,4180\,K, \logg{}\,=\,4.62\,dex, and [Fe/H]\,=\,$-0.69$\,dex based on low-resolution spectroscopic analysis. By fitting PHOENIX stellar models to high-resolution CARMENES spectra, \citet{2019A&A...627A.161P} reported \teff{}\,=\,3445,K, \logg{}\,=\,4.73\,dex, and [Fe/H]\,=\,$-0.05$\,dex. Similarly, \citet{2019A&A...625A..68S} estimated \teff{}\,=\,3513\,K, \logg{}\,=\,4.91\,dex, and [Fe/H]\,=\,$-0.03$\,dex based on high-resolution spectral analysis. \citet{2019AJ....158...87D}, using medium-resolution spectra, found \teff{}\,=\,3479\,K and [Fe/H]\,=\,0.07\,dex. \citet{2021A&A...656A.162M} and \citet{2021MNRAS.506..150B}, both using high-resolution spectral analysis, reported \teff{} values of 3563\,K and 3492\,K, \logg{} values of 4.93\,dex and 4.8057\,dex, and [Fe/H]\,=\,$-0.13$\,dex. Our estimated \teff{} is consistent with the spectroscopic estimates of \citet{2019A&A...627A.161P}, \citet{2019AJ....158...87D}, and \citet{2021A&A...656A.162M}. The metallicity value we report also aligns with those of \citet{2019A&A...627A.161P} and \citet{2019AJ....158...87D}. However, our estimated \logg{} is slightly higher than the values reported in the referenced literature.

\subsection{Mass-Radius relation and the Fulton Gap}\label{2.2}
Observational studies of close-in Kepler exoplanets (Period $P\,\leq\,100$ days) shows a bimodal distribution near 1.5\,-\,2\,$R_{\oplus}$ \citep{fulton2017california,fulton2018california,hardegree2020scaling}. This radius valley, usually called the Fulton Gap, separates the rocky super-Earth population from the gaseous rich mini-Neptunes and sub-Neptunes. This compositional transition between exoplanets, originating from similar formation mechanism, is believed to be period-dependent \citep{van2018asteroseismic,martinez2019spectroscopic} and also varies with the stellar properties, such as metallicity \citep{owen2018metallicity}, mass \citep{macdonald2019examining,cloutier2020evolution} and age \citep{berger2020gaia,david2021evolution}. The existence of this valley was primarily attributed to photoevaportion of hydrogen-rich exoplanet envelopes driven by energetic XUV rays from their host star in the initial 100Myr, when the star is most active
\citep{owen2013kepler,lopez2014understanding,lopez2018formation}. Another explanation is the core-powered mass loss mechanism, where the core of the planet release thermal energy, accumulated during formation, and cools down slowly in a timescale of Gyr \citep{ginzburg2018core,gupta2019sculpting,gupta2020signatures}. Most recently, the gas-poor accretion mechanism \citep{2014ApJ...797...95L, lee2021primordial} has also stand out as a possible technique for sculpting the radius valley, in which rocky super-Earths are supposed to form at later times when all the gas from the protoplanetary disk has dissipated. 

\begin{figure*}
    \includegraphics[width=0.7\linewidth]{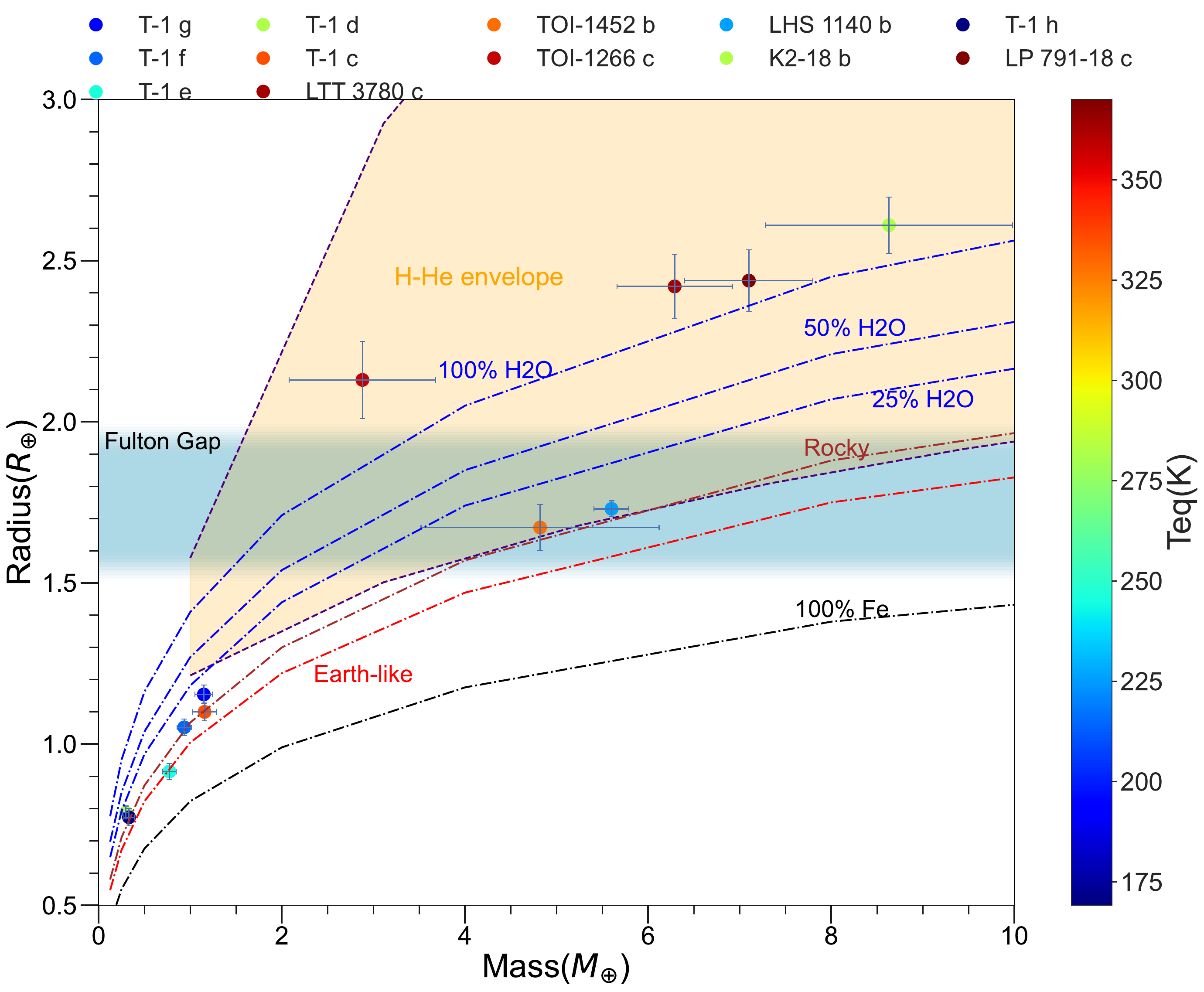}
    \caption{Mass-radius relationship for the 12 exoplanets, color-coded by their equilibrium temperature, as indicated by the colorbar. The compositional curves are taken from \citet{zeng2016mass}, while the H-He-rich region from \citet{rogers2023conclusive} is highlighted in orange. The blue-shaded area represents the radius valley, also known as the Fulton gap.}
    \label{fig:3}
\end{figure*}

Fig.~\ref{fig:3} displays the radius-mass distribution of the sample exoplanets, color coded according to their equilibrium temperature, with the horizontal shaded region, labeled as the Fulton Gap, indicating the radius valley between $1.5\,R_{\oplus}$ and $2R_{\oplus}$. The rest of the curves shown in Fig.~\ref{fig:3} represent the empirical mass-radius relations of \citet{zeng2016mass}. All the different compositional curves are labeled along with the H-He enveloped (0.1\%\,-\,30\%) exoplanet population space of \citet{rogers2023conclusive}, highlighted in orange. This region also marks the degeneracy where the exoplanet composition, either H-He or water-rich, of sub-Neptunes cannot be reliably determined from mass and radius alone. The empirical mass-radius relations \citep{dressing2015mass, zeng2017simple, zeng2016mass, 2019PNAS..116.9723Z} provide insight into the possible structural composition of exoplanets with varying mass and radius. While these relations do not offer exact predictions for planetary composition, they serve as a foundational tool for analyzing the structural and atmospheric characteristics of habitable exoplanets. The TRAPPIST-1 planets (denoted as T-1 in Fig.~\ref{fig:3}) lie along and between the rocky and Earth-like compositional curves, indicating their rocky or Earth-analogous composition. LHS 1140 b and TOI-1452 b are positioned within the gas envelope region, near the pure rocky composition curve. This placement suggests that these two planets could be volatile-rich, with a water mass fraction (WMF) of $\lesssim\,10\%$. The remaining four planets—TOI-1266 c, LTT 3780 c, LP 791-18 c, and K2-18 b—are located in the water-rich region, indicating the potential for H-He envelopes. Sec.~\ref{2.3} provides a detailed structural and compositional analysis of all the exoplanets, excluding the TRAPPIST-1 system.

As previously mentioned, the radius valley serves to distinguish two distinct exoplanet populations based on their composition: sub-Neptunes and super-Earths. Exoplanets within the radius valley are prime candidates for studying planetary formation mechanisms. Observing the long-term evolution of these planets can offer valuable insights into the processes that may have shaped the radius valley and the factors influencing the divergent pathways of their development. Each of the formation mechanisms discussed above leads to distinct planetary radius-period relations \citep{owen2017evaporation, lopez2018formation, gupta2019sculpting, 2020A&A...638A..52M, lee2021primordial}, which contribute to the formation of the radius valley around main-sequence stars. Recently, \citet[hereafter M19]{martinez2019spectroscopic}, based on the occurrence rate of close-in small exoplanets around Sun-like stars, suggested a radius-period relation 
\begin{eqnarray}\label{eq:eq_m19}
\left( \frac{d \log R_\textrm{valley}}{d \log P} \right)_{\textrm{M19}}=-0.11\pm0.02,
\end{eqnarray}
which is consistent with all the thermally driven mass loss (TDML) mechanisms. Later, \citet[hereafter CM20]{cloutier2020evolution} demonstrated that, if super-Earths were formed solely by accumulating planetary embryos in a gas-poor environment, commonly referred to as gas-depleted formation (GDF), then the radius-period relation would have a positive slope, given by: 
\begin{eqnarray}\label{eq:eq_cm20}
\left( \frac{d \log R_\textrm{valley}}{d \log P} \right)_{\textrm{CM20}}=0.058\pm0.022.
\end{eqnarray}

These calculations were specifically conducted for low-mass stars (\teff\,$\leq 4700\,K$), suggesting that gas-depleted super-Earth formation may be particularly dominant around M-dwarfs.\\

\begin{figure}
    \includegraphics[width=1.1\linewidth]{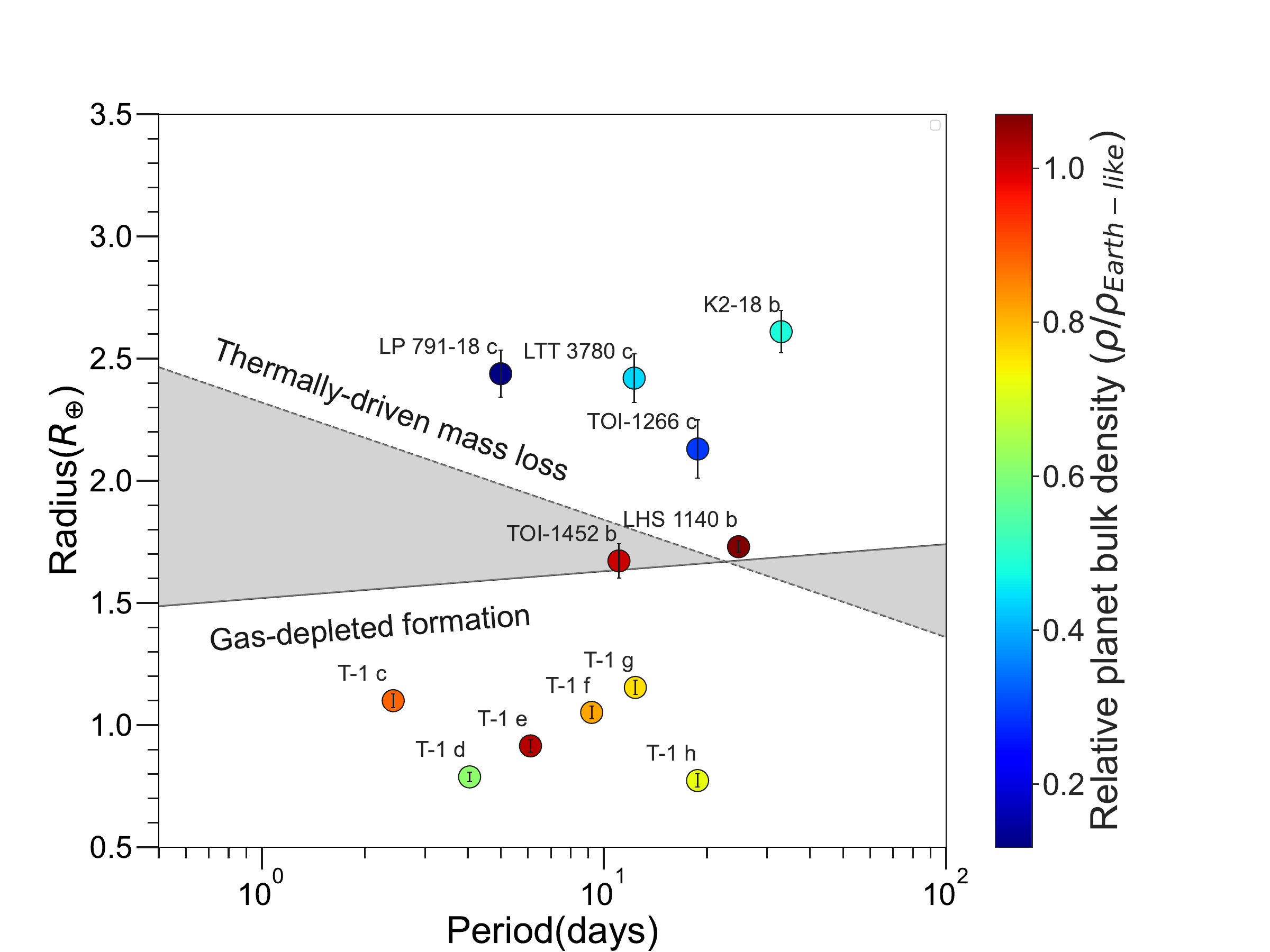}
    \caption{Radius-period relationship of exoplanets. The exoplanets in the sample are color-coded according to their bulk densities. The black solid lines depict the radius-period trends for the TDML and GDF formation mechanisms, as defined by M19 and CM20, respectively. The gray-shaded region highlights the overlap between the two trends.}
    \label{fig:4}
\end{figure}

Fig.~\ref{fig:4} presents the models of M19 and CM20, which predict the bulk composition of planets forming through the two different mechanisms and delineate different regions in the radius-period parameter space \citep{cadieux2022toi}. The stellar mass in the M19 model has been scaled down to match the mass distribution of CM20, following Eq.~11 therein. Our target exoplanets are overlaid on the predicted models. The exoplanets are color-coded according to their respective bulk densities relative to an Earth-like composition. Lighter, potentially water-rich planets are situated well above the shaded region, while the denser, terrestrial Trappist-1 planets are scattered below it.\\

The two super-Earths, LHS 1140 b and TOI-1452 b, with radius $1.73R_{\oplus}$ \citep{cadieux2024new} and $1.672R_{\oplus}$ \citep{cadieux2022toi}, respectively fall within the Fulton gap and serve as excellent targets to study the formation and evolution of planets within this radius valley. The locus of LHS 1140 b in fig.~\ref{fig:4} lies close to both the formation mechanisms, indicating that either pathway could be a dominant process responsible for the formation of this planet. For TOI-1452 b, although its locus shows inconsistency with the TDML model, the accurate position in the radius-period space and the observed bulk density cannot fully support the GDF model as well \citep{cadieux2022toi}. A detailed discussion on evolution of atmospheric envelope of these planets in section~\ref{3}, hints towards a mini-Neptune like composition with very small amount of H-He layer above their cores. These small amount of gaseous envelope can be either due to their gas poor formation or the loss of atmosphere due to stellar erosion. Therefore, solely based on the observed planetary mass, density or present gaseous envelope it is difficult to support either of the formation models as the dominant mechanism for the emergence of the radius valley.

\subsection{Interior Structure Composition}\label{2.3}
Fig.~\ref{fig:3} presents a schematic of the possible composition of exoplanets within the H-He envelope region. In this section, we model the interior composition of these exoplanets using the \textcolor{gray}{\texttt{ExoMDN}}\footnote{\href{https://github.com/philippbaumeister/ExoMDN.git}{ExoMDN}} framework \citep{2023A&A...676A.106B}. This framework is built on a machine learning algorithm that infers the interior structure of exoplanets based on their mass, radius, and equilibrium temperature. The model was trained on a dataset of 5.6 million synthetic interior structures consisting of a core, mantle, water and ice layers, and an H-He envelope. \textcolor{gray}{\texttt{ExoMDN}} utilizes Mixture Density Networks (MDN) \citep{bishop1994mixture}, which provide a probability distribution for the various layers through a linear combination of Gaussian kernels. Later \citet{2023A&A...676A.106B} showed that mass, radius, and equilibrium temperature alone are insufficient to precisely constrain the interior structure of exoplanets. To improve the modeling, they incorporated the fluid Love number $k_2$ \citep{love1909yielding} as a fourth parameter. The value of $k_2$ is influenced by the planet's interior density distribution and plays a crucial role in determining the shape of a rotating exoplanet in hydrostatic equilibrium \citep{padovan2018matrix, kellermann2018interior, baumeister2020machine}. The measurement of $k_2$ requires the estimation of orbital's apsidal precession \citep{csizmadia2019estimate} and second-order measurements of the transit light curve shape \citep{akinsanmi2024tidal, barros2022detection, hellard2019retrieval}. These high precession measurements obscure the precise retrieval of love number value for each exoplanet. Consequently, a common assumption is to adopt the Love number of Earth as $k_2 = 0.933$ \citep{lambeck1980estimates} and for Neptune as $k_2 = 0.392$ \citep{2023MNRAS.522.4251F} for planets with densities similar to these planets. Based on Fig.~\ref{fig:3} and Table~\ref{tab1}, we assume $k_2=0.933$ for LHS 1140 b and TOI-1452 b (having densities similar to Earth) and $k_2=0.392$ for rest of the four planets, TOI-1266 c, LP 791-18 c, LTT 3780 c and K2-18 b (having densities similar and close enough to Neptune, i.e., $0.297 \rho_{\oplus}$\footnote{\href{https://nssdc.gsfc.nasa.gov/planetary/factsheet/}{NASA Planetary factsheet}}).

\begin{figure*}
    \includegraphics[width=0.33\linewidth]{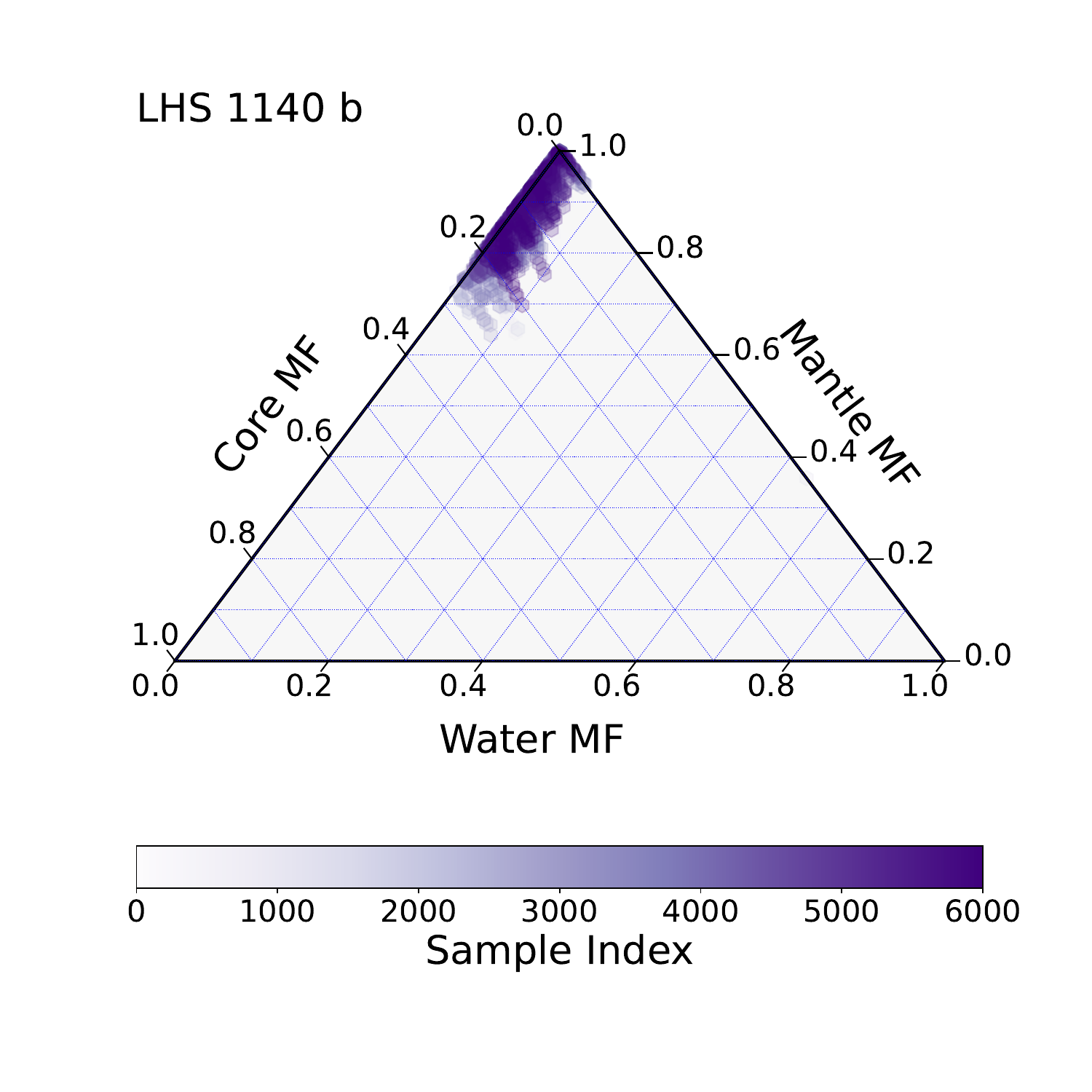}
    \includegraphics[width=0.33\linewidth]{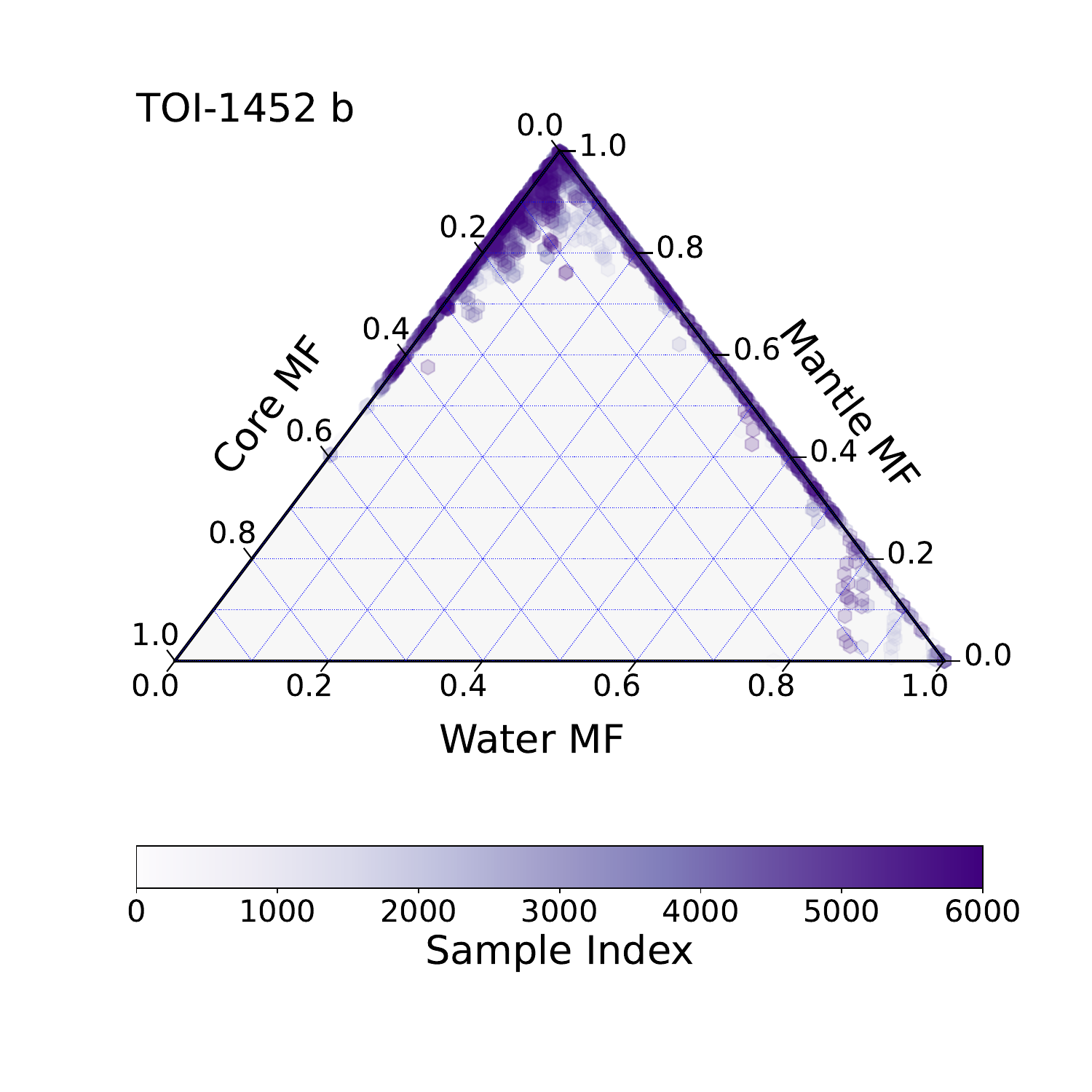}
    \includegraphics[width=0.33\linewidth]{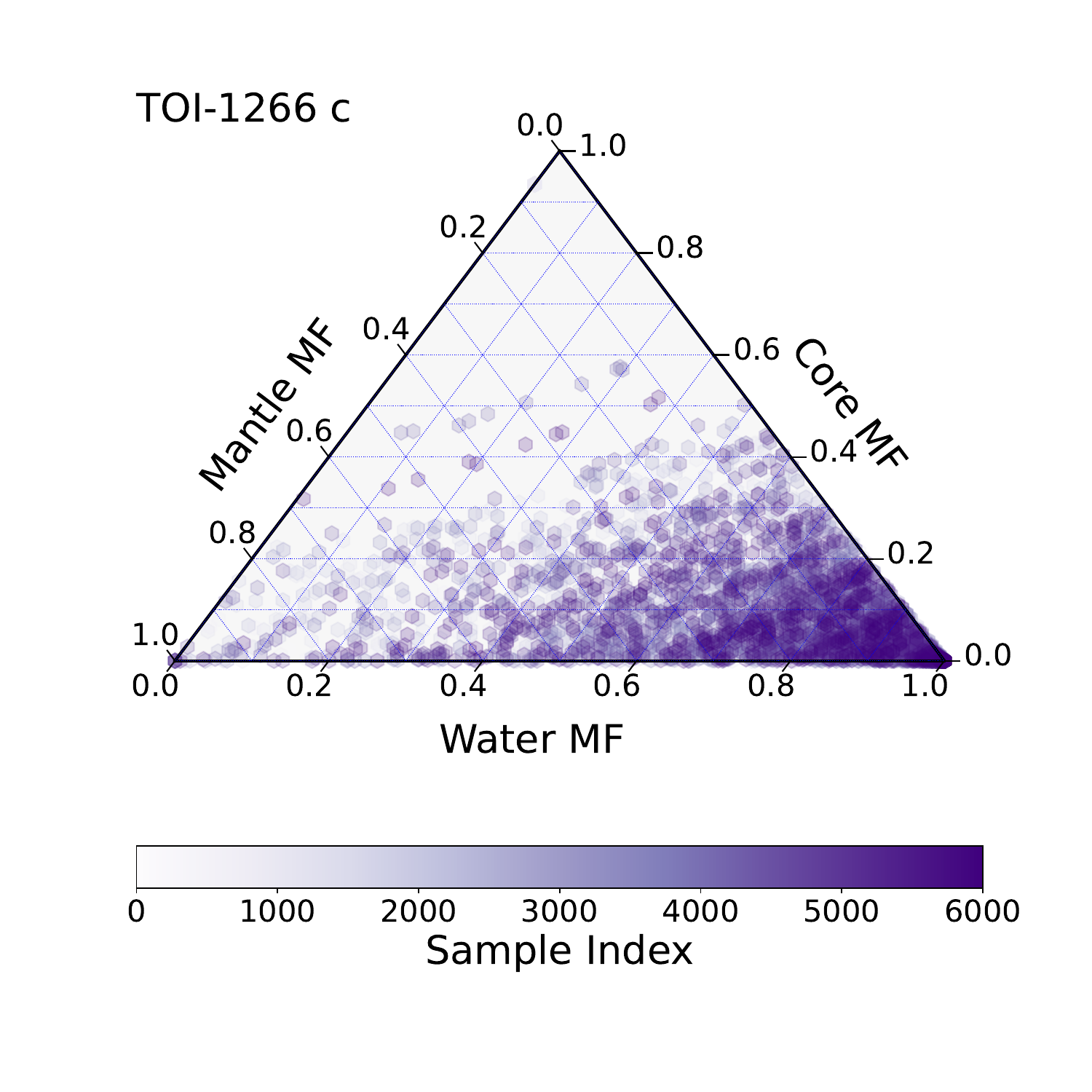}
    \includegraphics[width=0.33\linewidth]{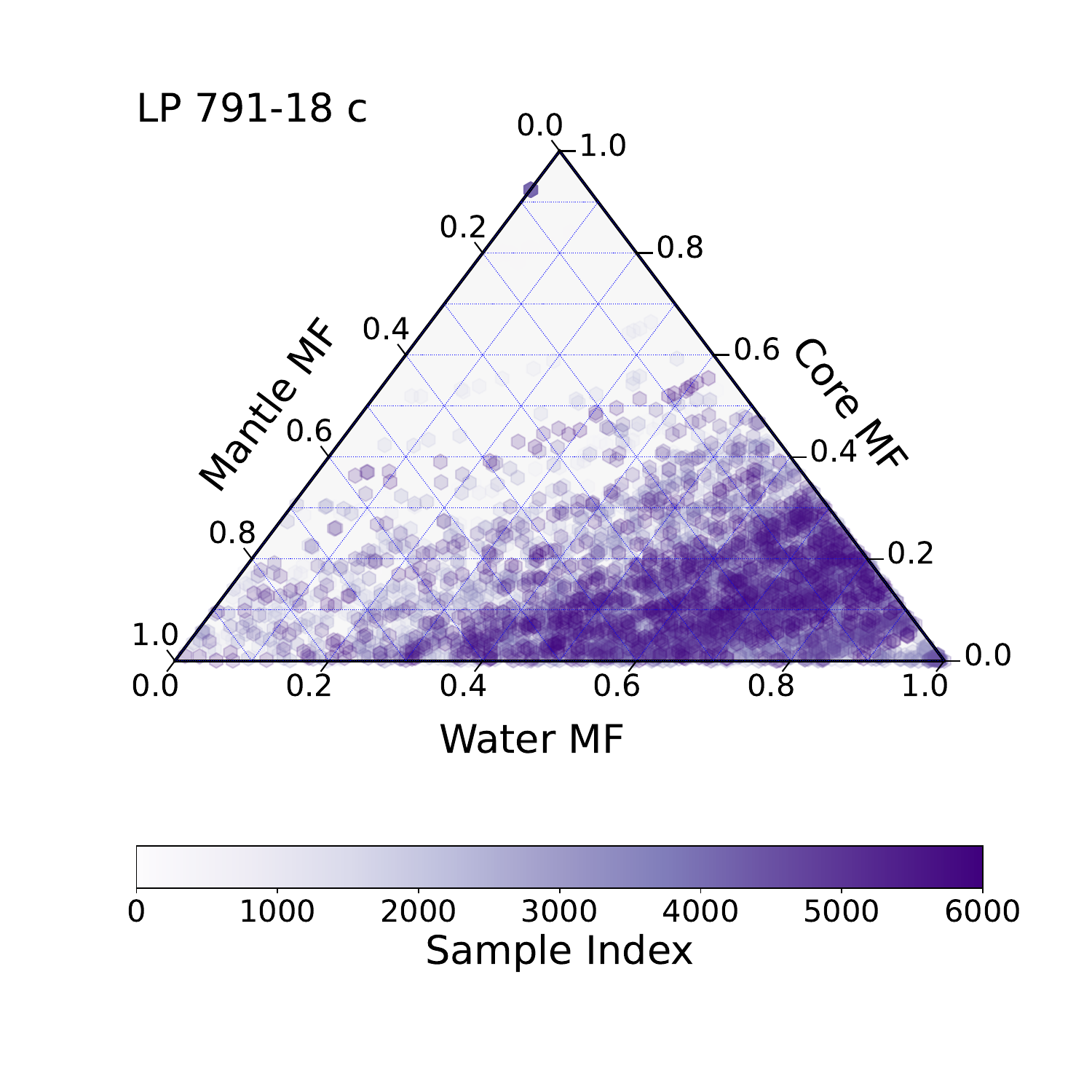}
    \includegraphics[width=0.33\linewidth]{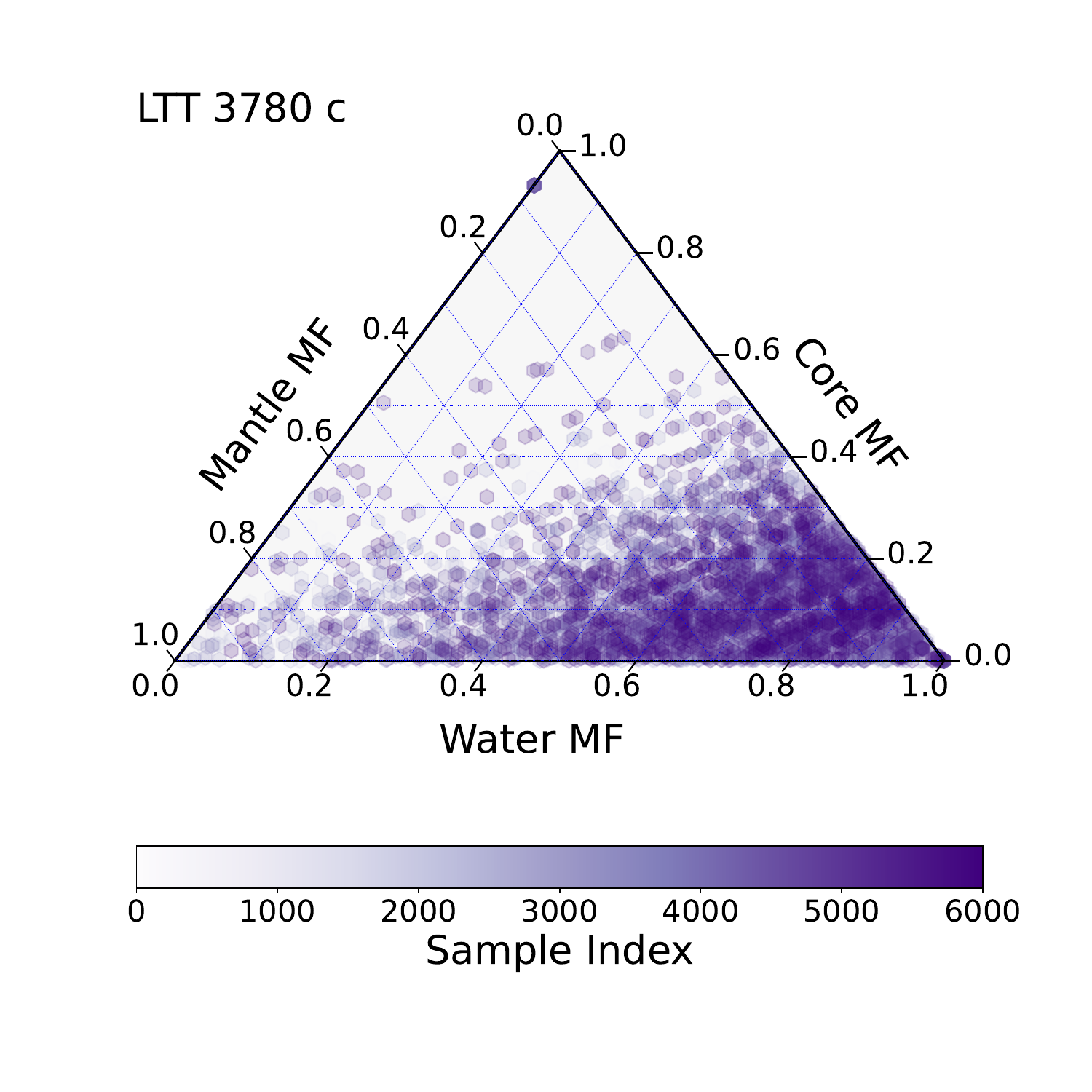}
    \includegraphics[width=0.33\linewidth]{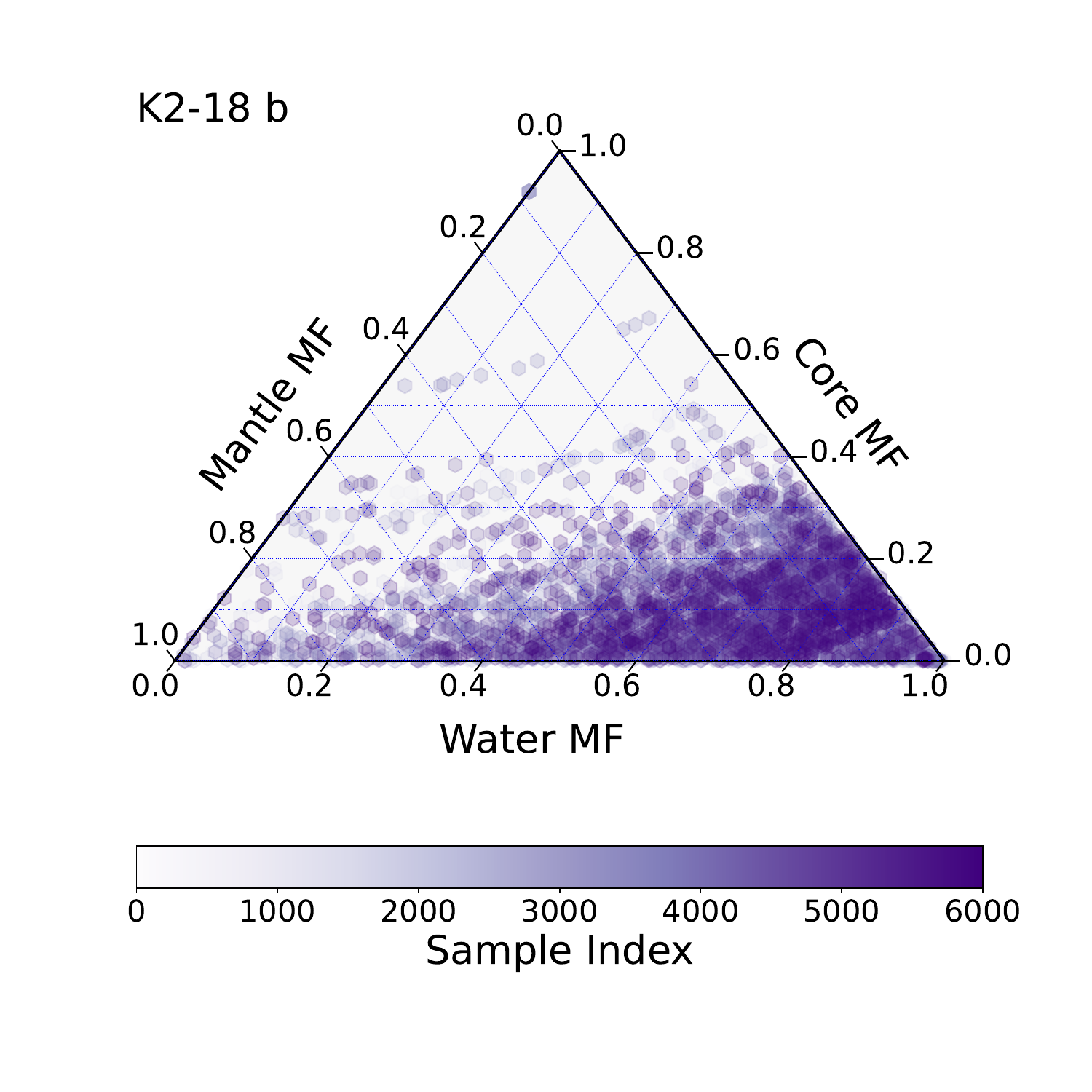}
    \caption{Ternary plots showing various composition of CMF, MMF and WMF for the sample exoplanets, obtained using \texttt{ExoMDN}. The scatter points shows the weightage of the different mass fractions. The high density points describes the most probable composition of the exoplanet }
    \label{fig:5}
\end{figure*}

We used the mass, radius, equilibrium temperature, and fluid love number as input parameters for \textcolor{gray}{\texttt{ExoMDN}}, incorporating uncertainties by drawing a normal distribution of each parameter with a standard deviation equal to their observed uncertainties. Following \citet{2023MNRAS.522.4251F}, we considered a $10\%$ uncertainty in the love numbers. Figs.~\ref{fig:5} and~\ref{fig:6} present the results of different mass fractions of the five exoplanets as ternary and histogram plots, respectively. In Section~\ref{3}, we present a detailed discussion on the interior structure of all the individual exoplanets.

\subsection{Hydrodynamic Atmospheric Escape}\label{2.4}
Exoplanets with radii greater than $1.6 R_{\oplus}$ are generally not rocky \citep{rogers2015most}, and are likely to have a volatile-rich composition, a H-He rich envelope, or a combination of both \citep{lozovsky2018threshold}. Exoplanets within the radius valley and the sub-Neptune region, capable of gas accretion, can sustain a gaseous envelope above their terrestrial or icy core \citep{rogers2023conclusive}. Some close-in super-Earths may retain a portion of their primordial H/He envelope even after undergoing core-powered mass loss and photoevaporation \citep{misener2021cool}. Assuming that these exoplanets were capable of accreting H-He gas envelopes during their formation, we assess the likelihood of our six target planets retaining their primordial envelopes after experiencing thermally driven mass loss. We also verify the agreement of the exoplanets with TDML formation mechanism.

We use the publicly available code \textcolor{gray}{\texttt{photoevolver}}\footnote{\href{https://github.com/jorgefz/photoevolver.git}{PHOTOEVOLVER}} \citep{2023MNRAS.522.4251F} to model the host star's XUV-driven photoevaporation and core-powered mass loss over 10 Gyr, and examine the current envelope mass fraction of the exoplanets.

As inputs to the code, we use the interior structure composition models from \citet{2019PNAS..116.9723Z}, the envelope structure model of \citet{owen2017evaporation}, and the core-powered mass loss model from \citet{gupta2019sculpting} to study the evolution of the envelope fraction on planets with varying interior compositions\footnote{For detailed information regarding the stellar evolution and various mass loss models, please refer to the GitHub repository.}. The other inputs required by the code include the core radius of the planet, the mass of the host star, the semi-major axis, and the orbital period of the planet. Using the interior composition model, the code estimates the core mass, total mass, envelope radius and total radius of the planet at each age interval, ranging from 1 Myr to 10 Gyr. The next section provides a detailed discussion on the atmospheric modeling of the exoplanets.

\begin{figure*}
    \centering
    \begin{minipage}{\linewidth}
        \includegraphics[width=0.25\linewidth]{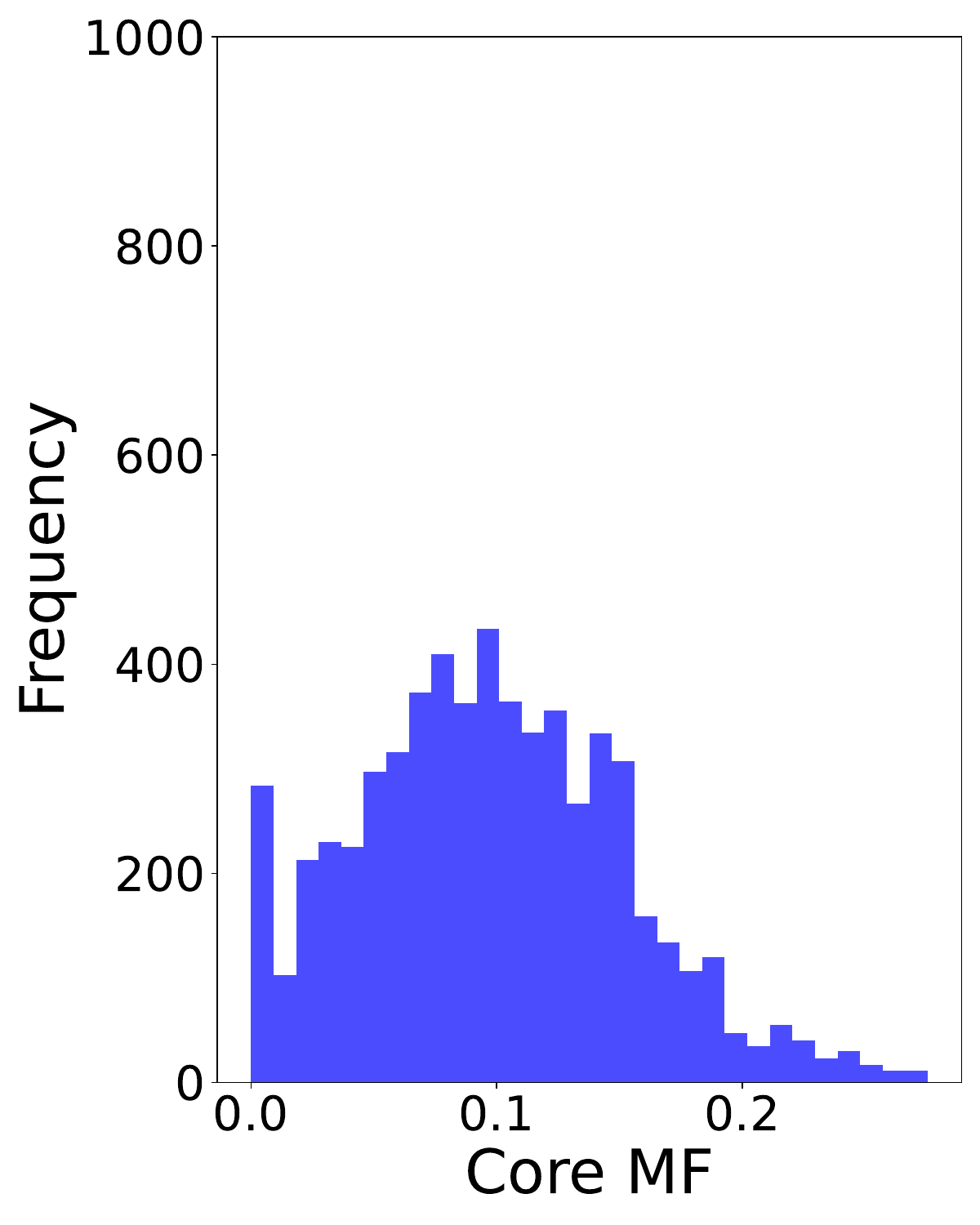}
        \includegraphics[width=0.25\linewidth]{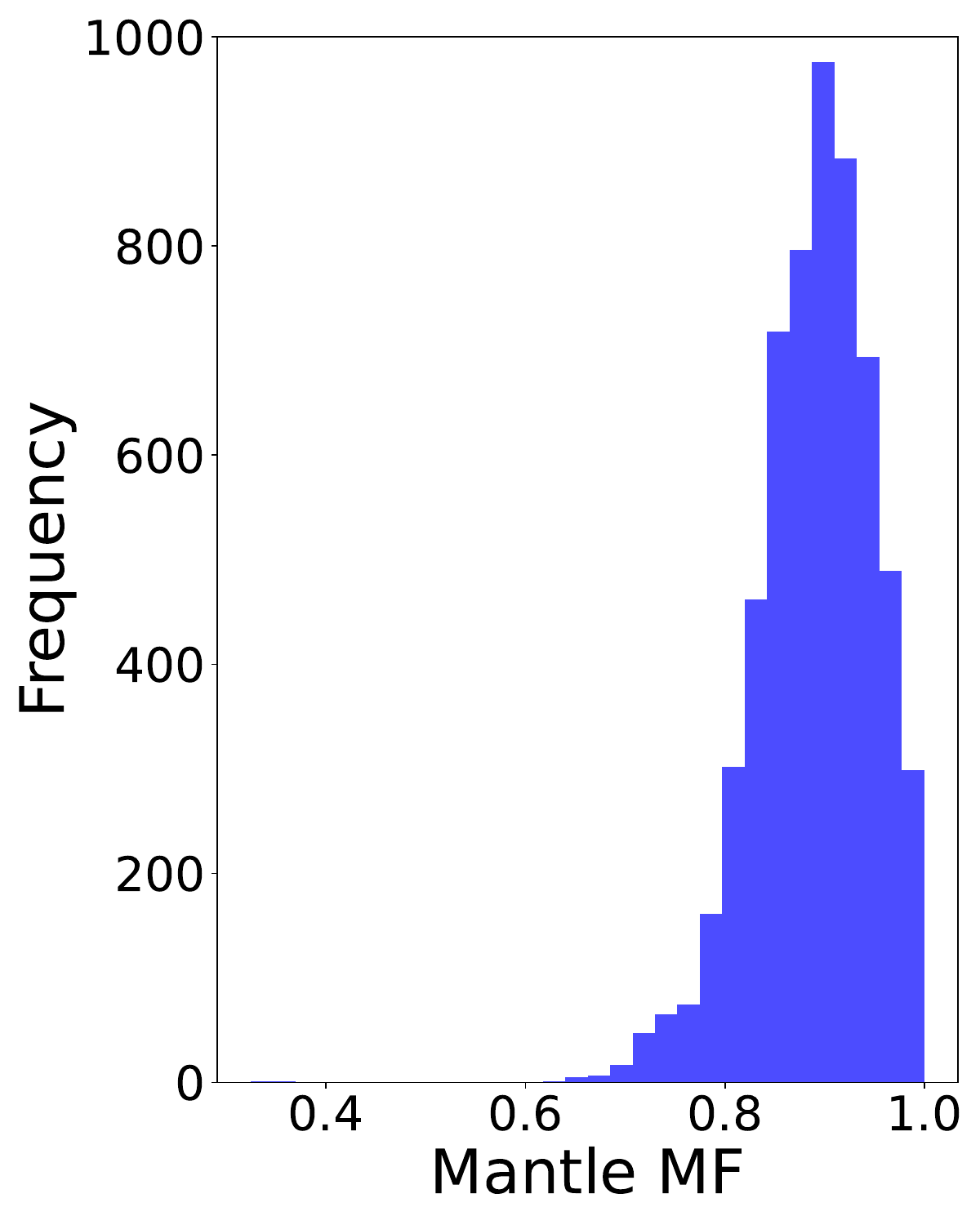}
        \includegraphics[width=0.25\linewidth]{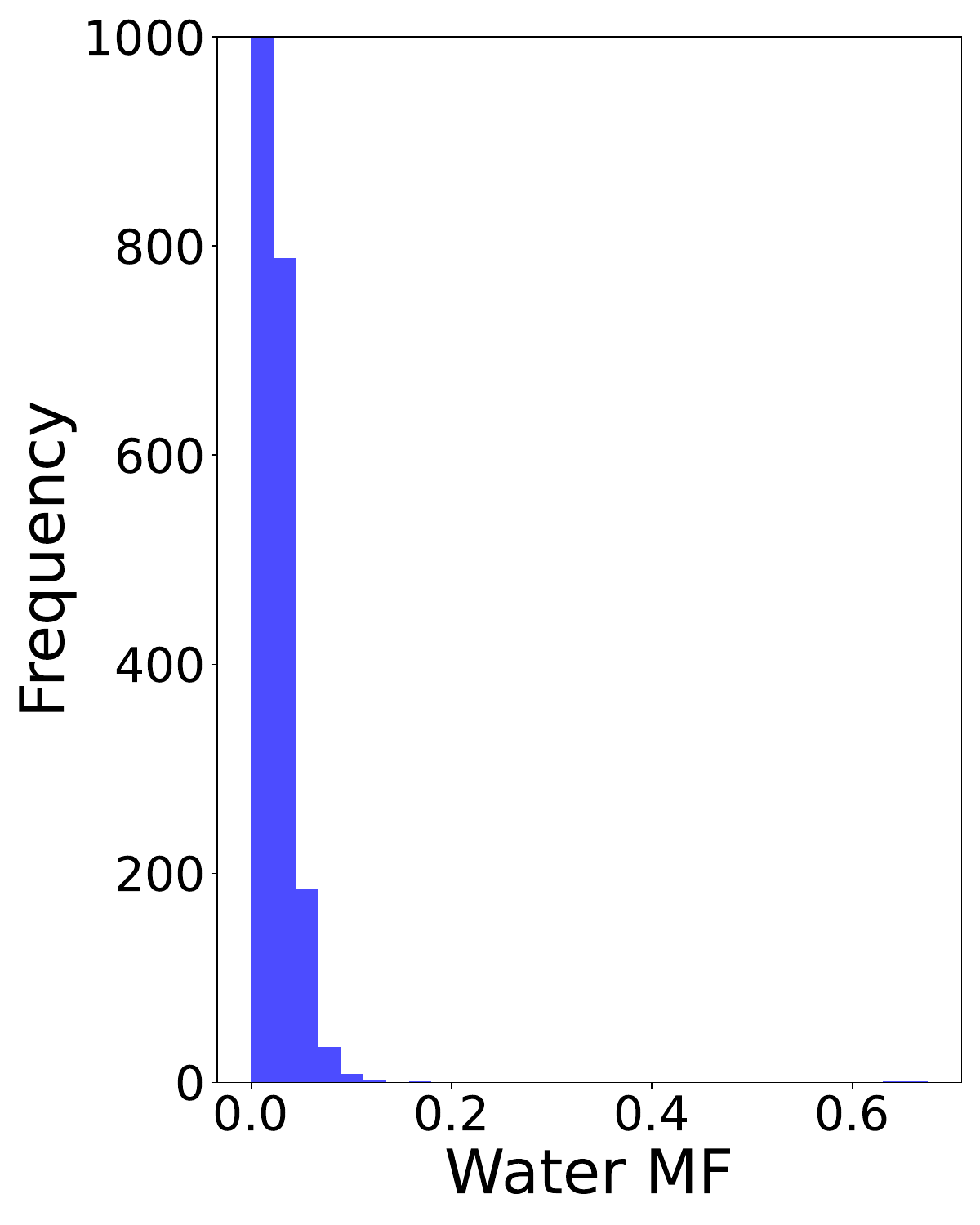}
        \includegraphics[width=0.25\linewidth]{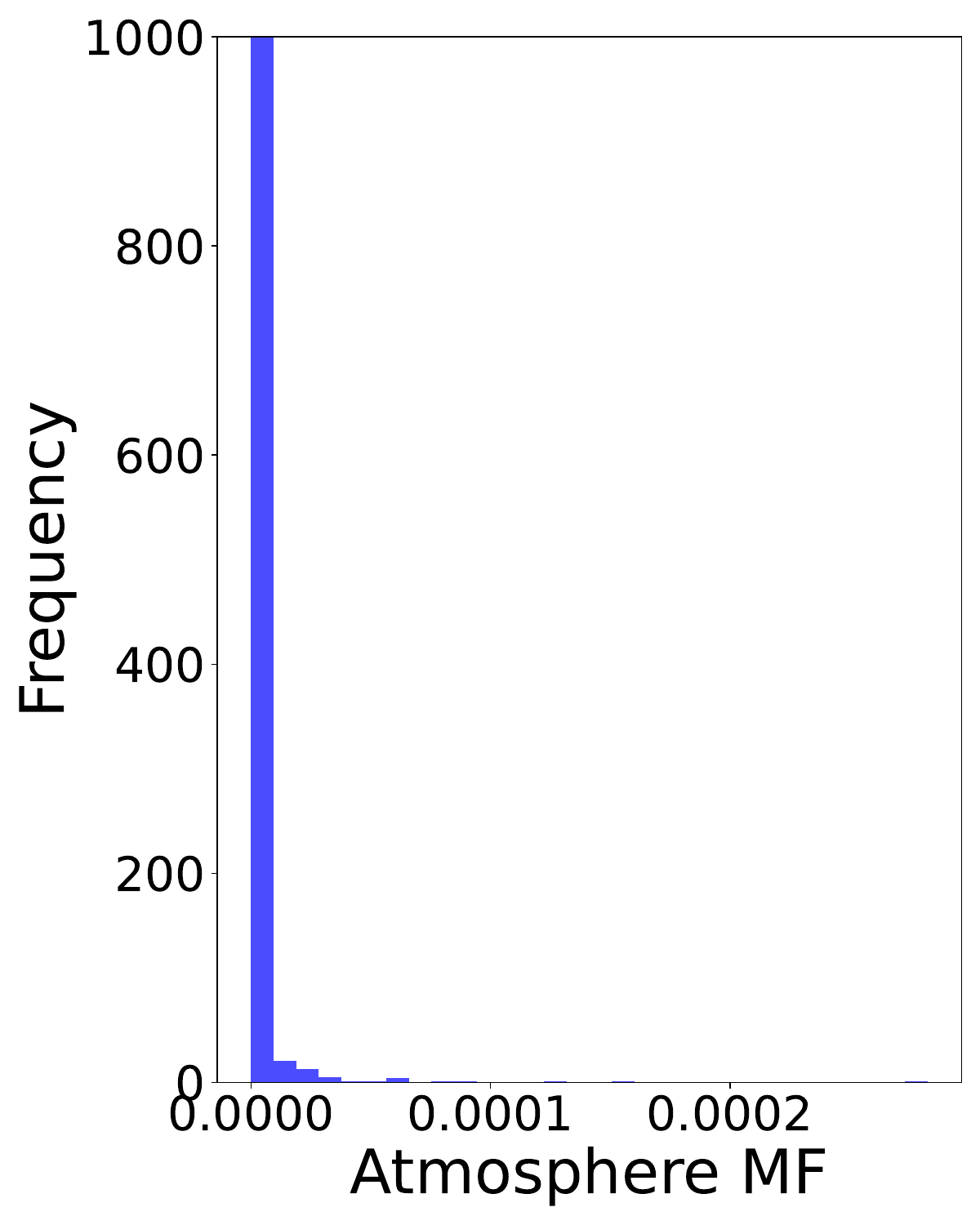}
        \subcaption*{LHS 1140 b}
    \end{minipage}
    
    \begin{minipage}{\linewidth}
        \includegraphics[width=0.25\linewidth]{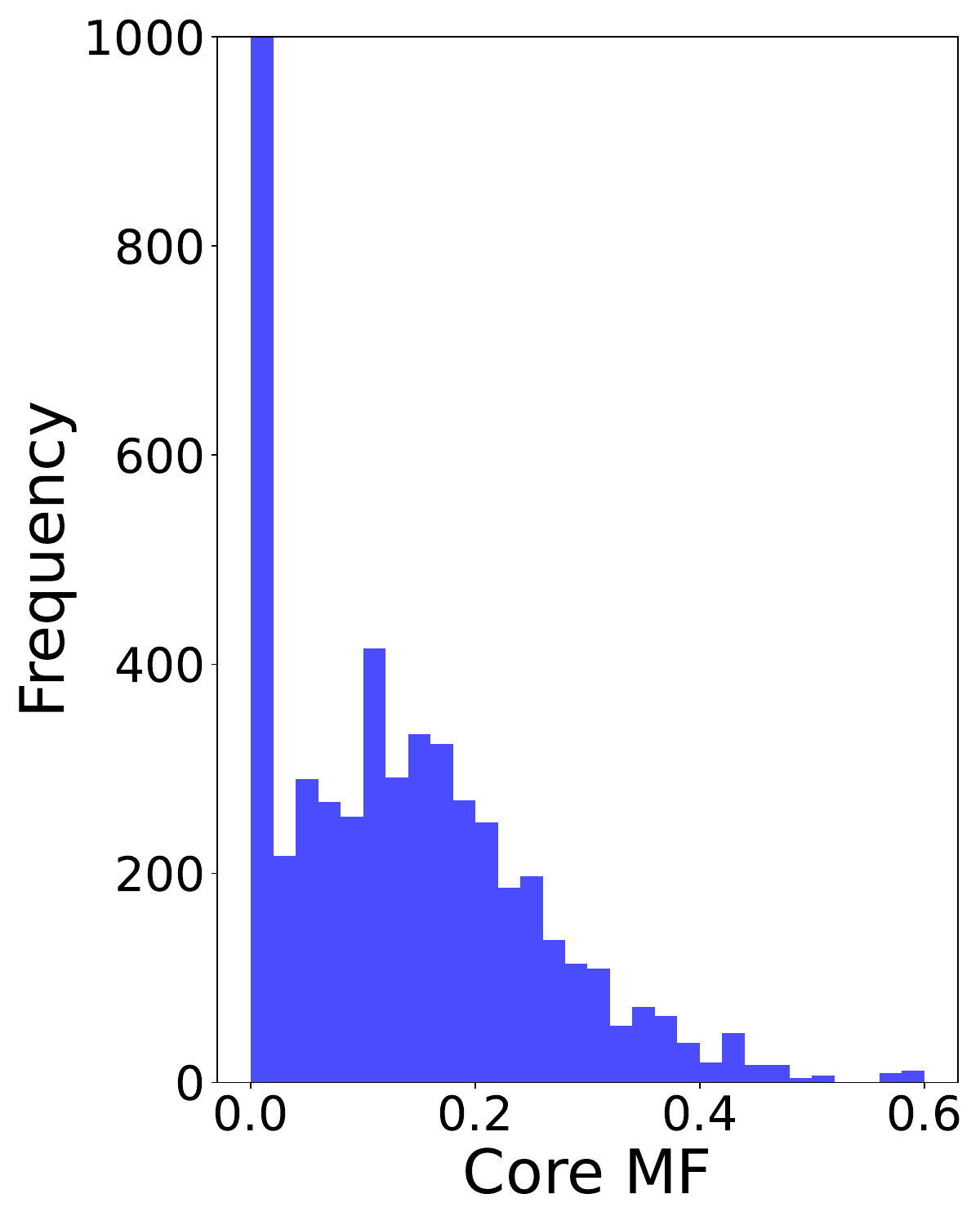}
        \includegraphics[width=0.25\linewidth]{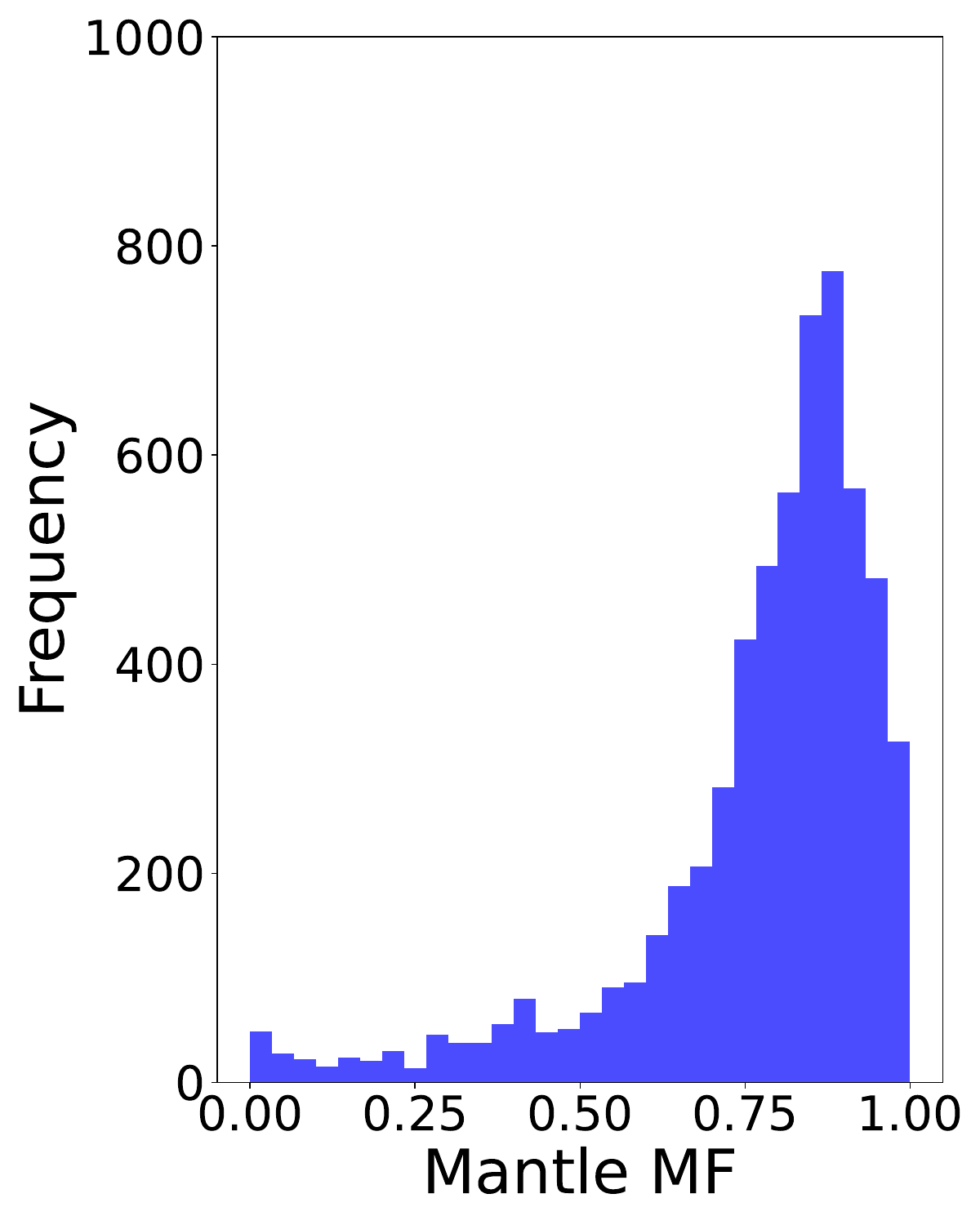}
        \includegraphics[width=0.25\linewidth]{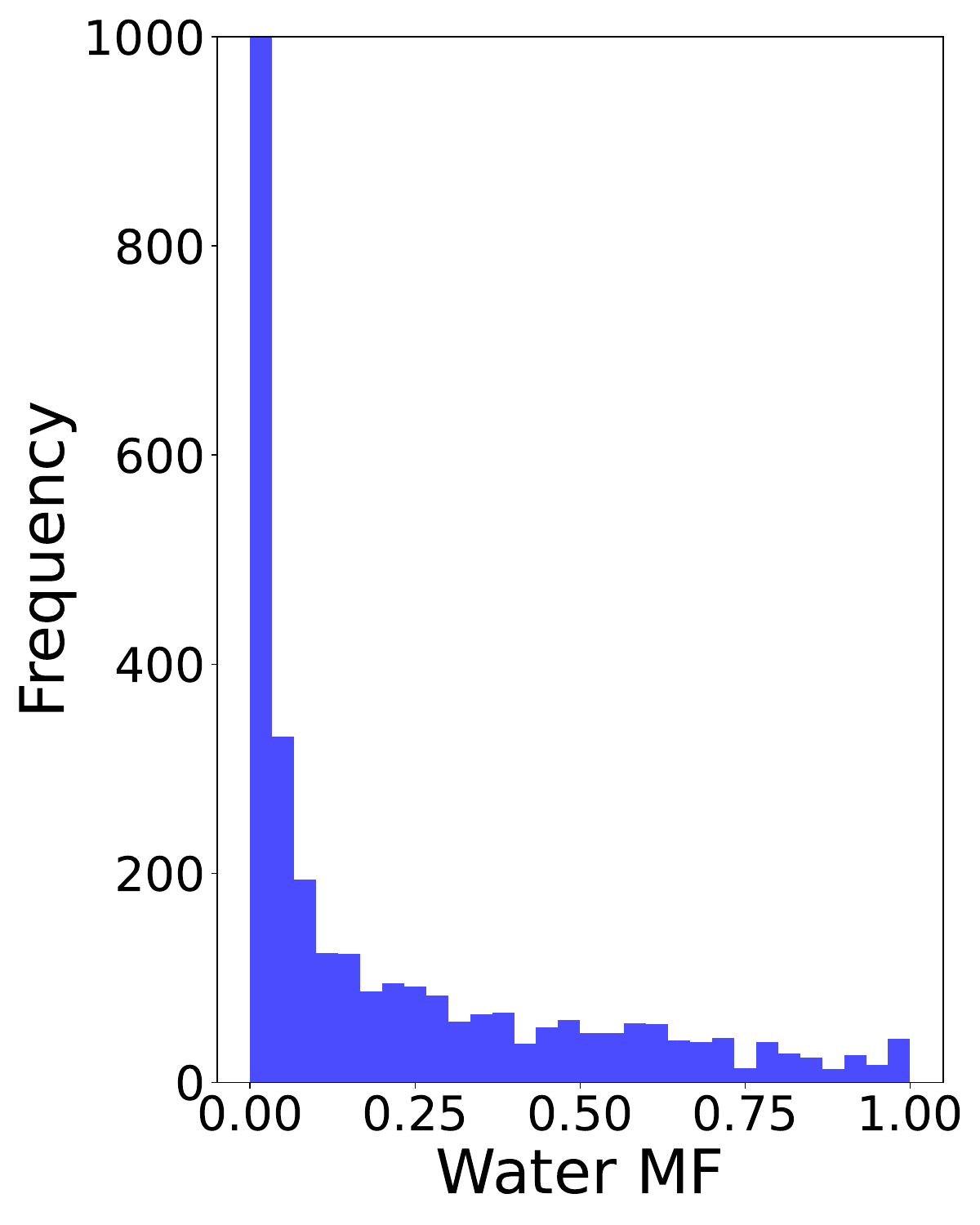}
        \includegraphics[width=0.25\linewidth]{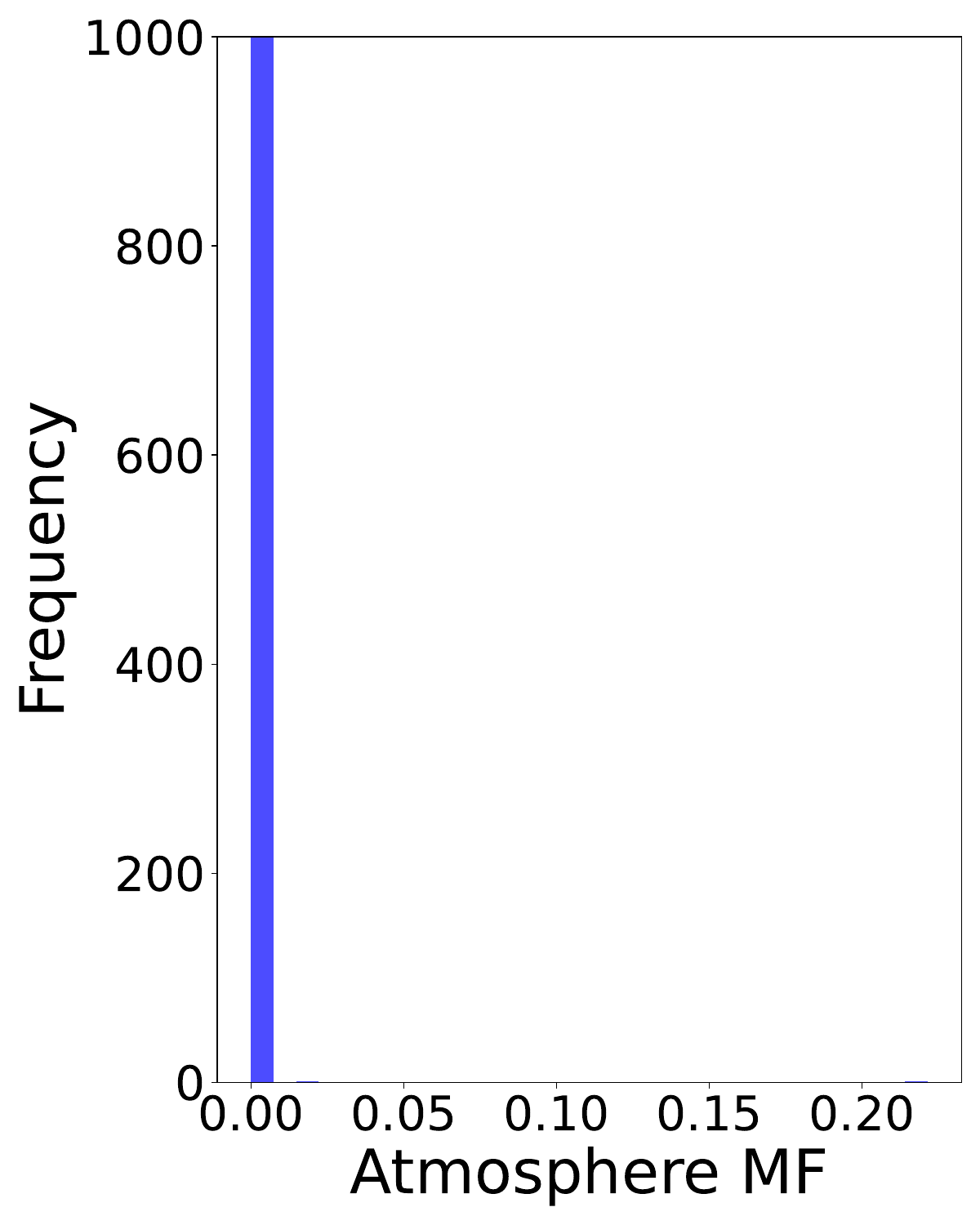}
        \subcaption*{TOI-1452 b}
    \end{minipage}

    \begin{minipage}{\linewidth}
        \includegraphics[width=0.25\linewidth]{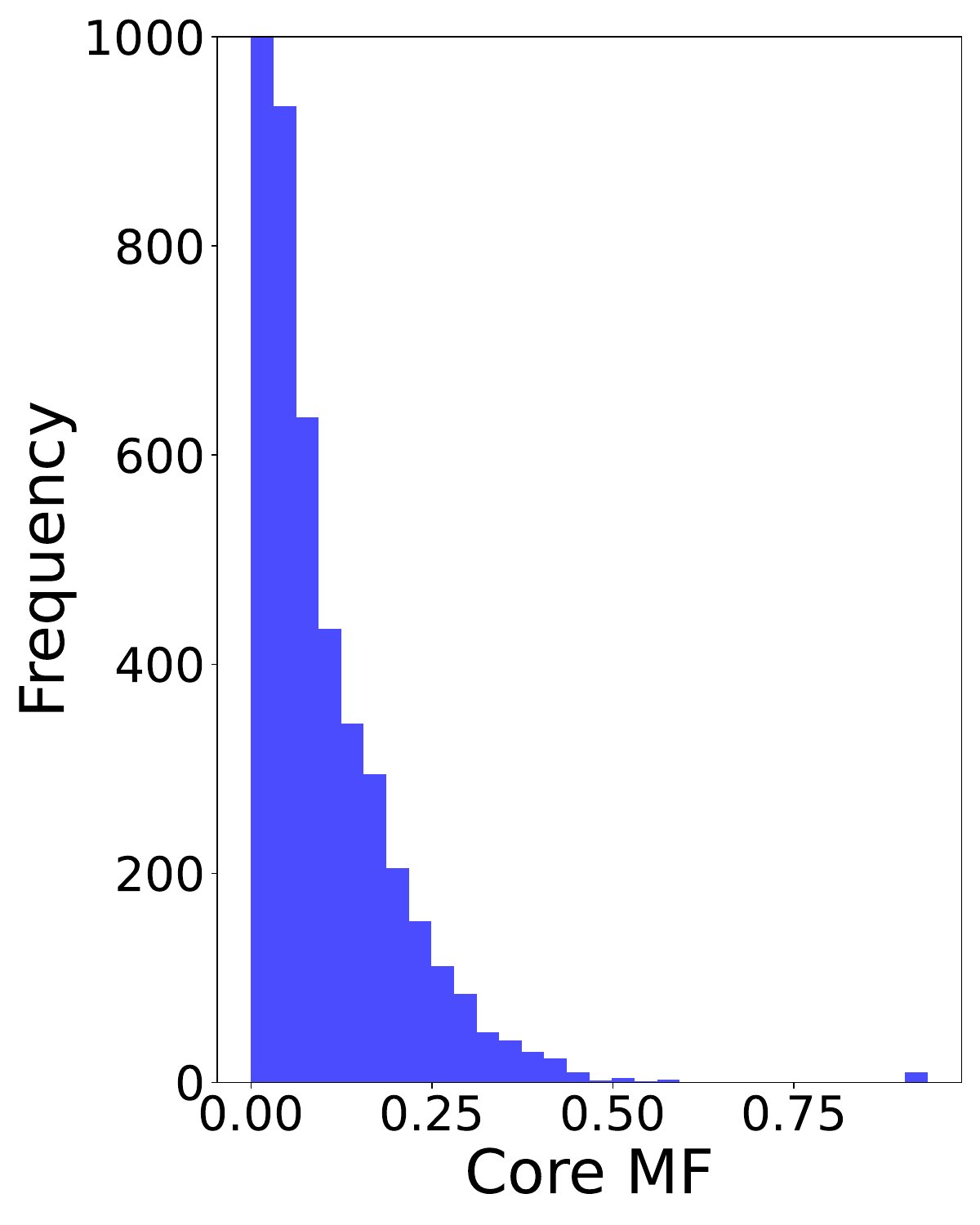}
        \includegraphics[width=0.25\linewidth]{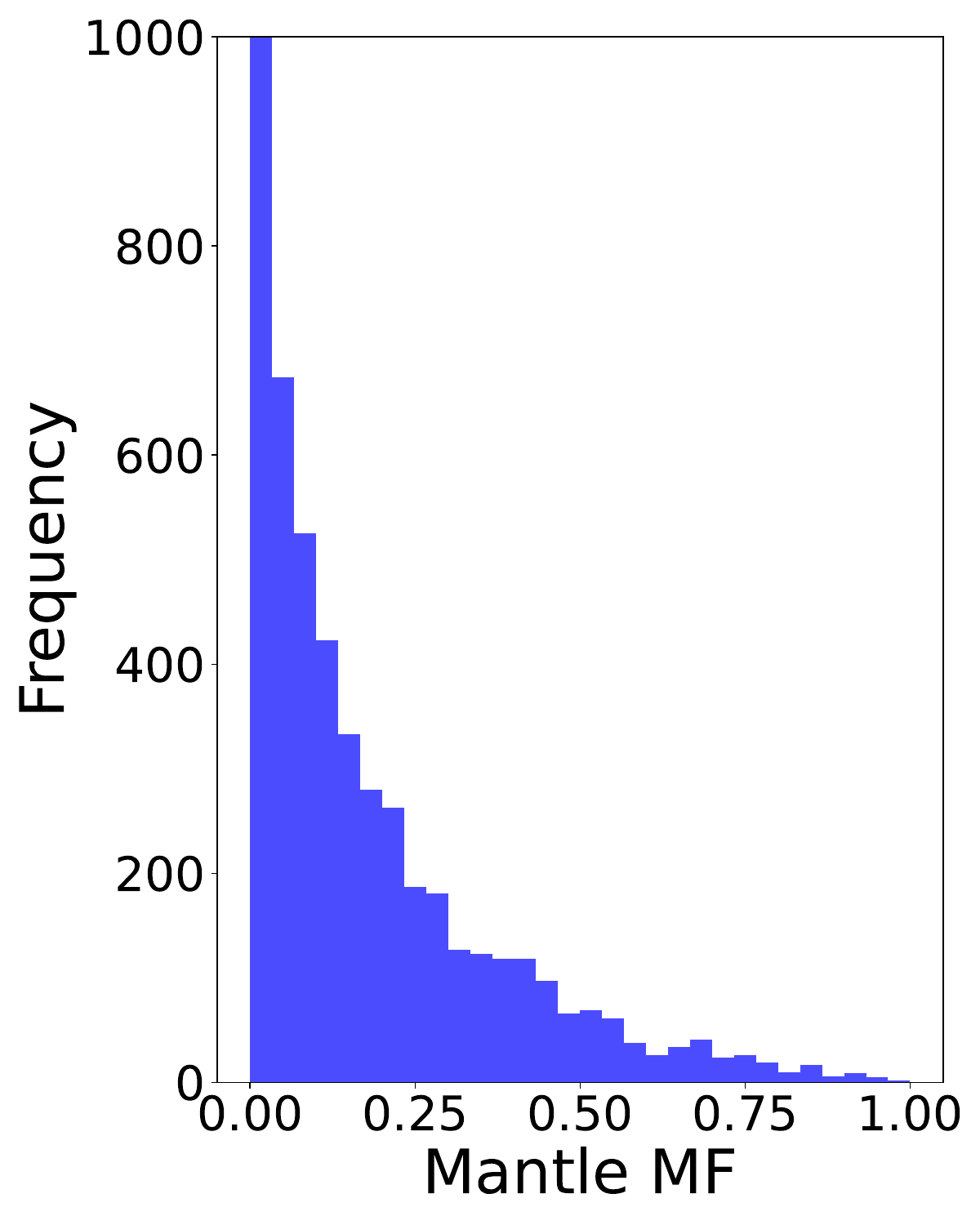}
        \includegraphics[width=0.25\linewidth]{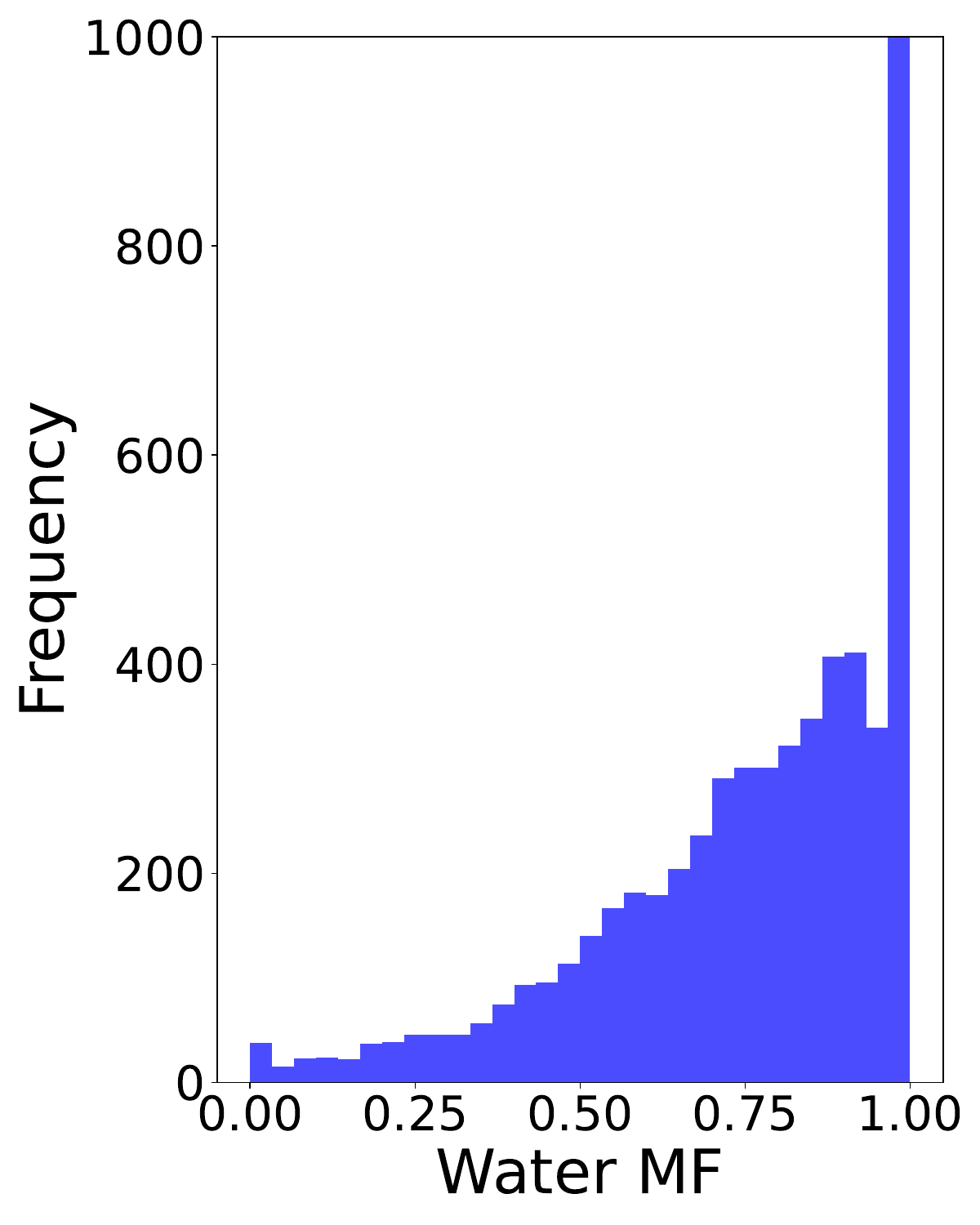}
        \includegraphics[width=0.25\linewidth]{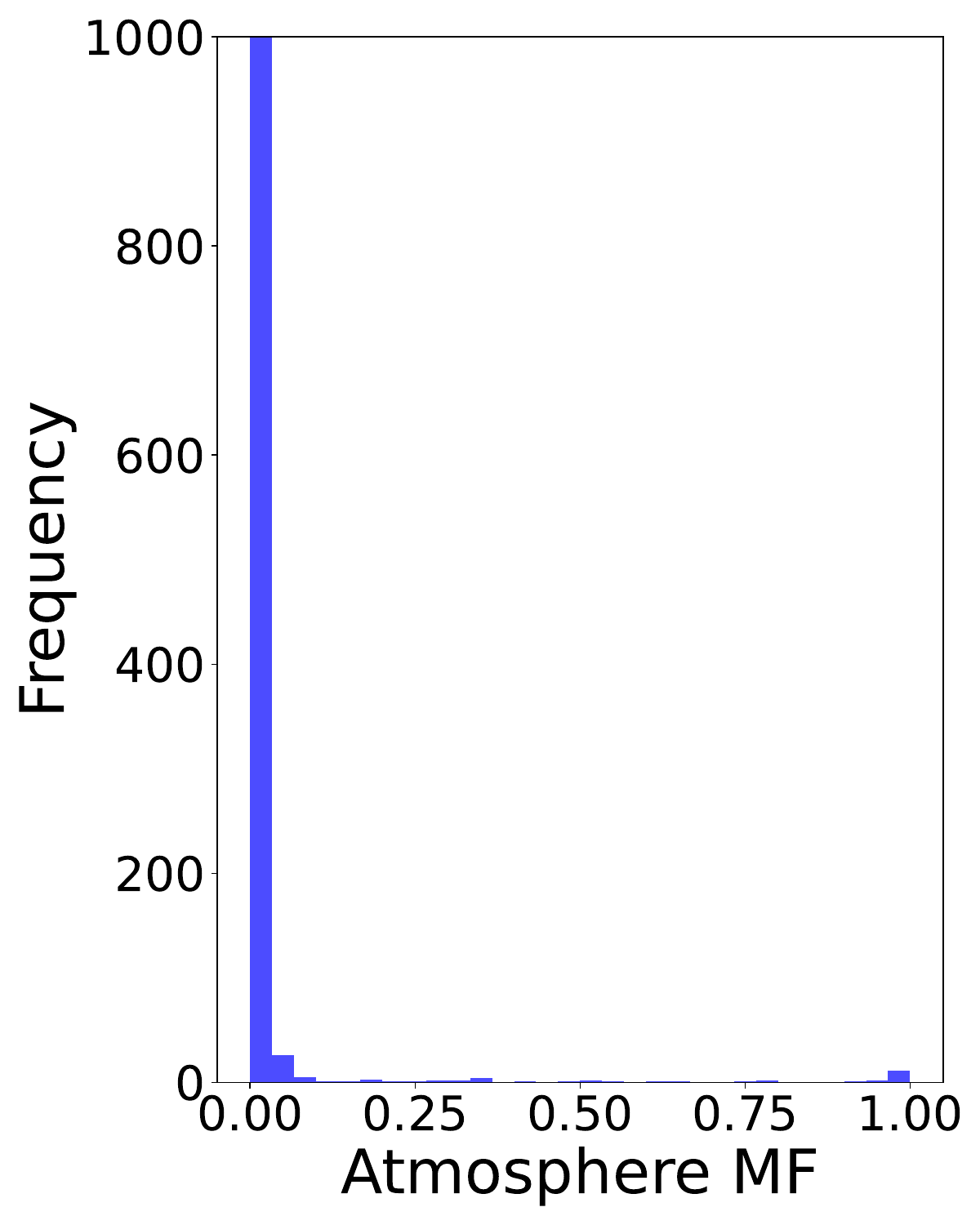}
        \subcaption*{TOI-1266 c}
    \end{minipage}
    
    \caption{Histograms for CMF, MMF, WMF, and AMF for individual exoplanets. The histograms are generated using the \texttt{ExoMDN}. LHS 1140 b and TOI-1452 b exhibits pridominantly rocky compositions, with MMF peaking around 80$\%$ for both the planets. While LHS 1140 b shows a negligible WMF, TOI-1452 b presents a very low-probability tail in the distribution, suggesting the potential for a WMF of up to 25$\%$. LP-791-18 c, LTT 3780 c and K2-18 b shows volatile-rich composition, characterized by WMF peaking around 70$\%$, and minimal contributions from the atmospheric mass fraction (AMF). Exoplanet TOI-1266 c exhibits primarily a water dominated composition, with a peak around 80$\%$ for the WMF. (1/2)}
    \label{fig:6}
\end{figure*}

\begin{figure*}
    \ContinuedFloat
    \centering

    \begin{minipage}{\linewidth}
        \includegraphics[width=0.25\linewidth]{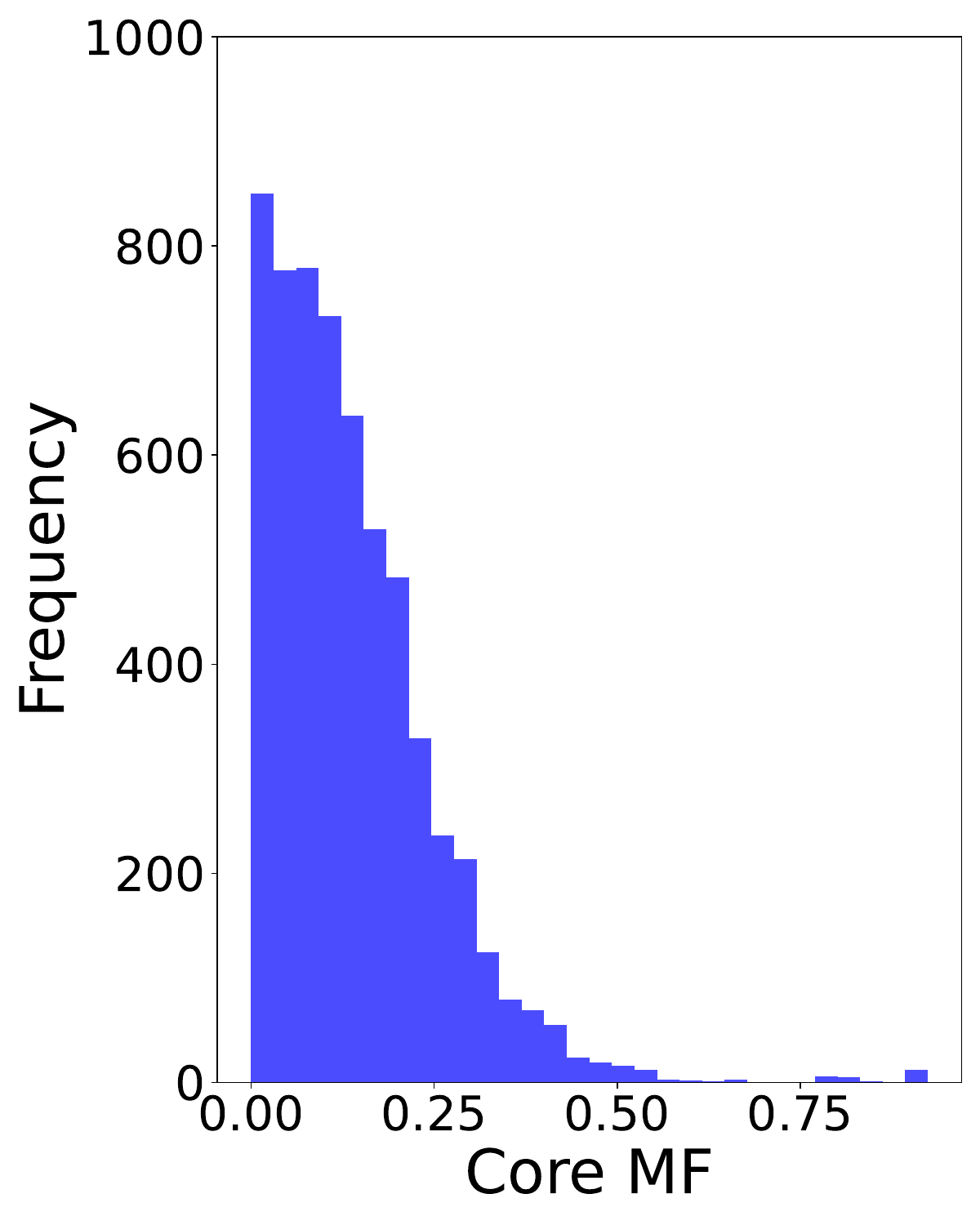}
        \includegraphics[width=0.25\linewidth]{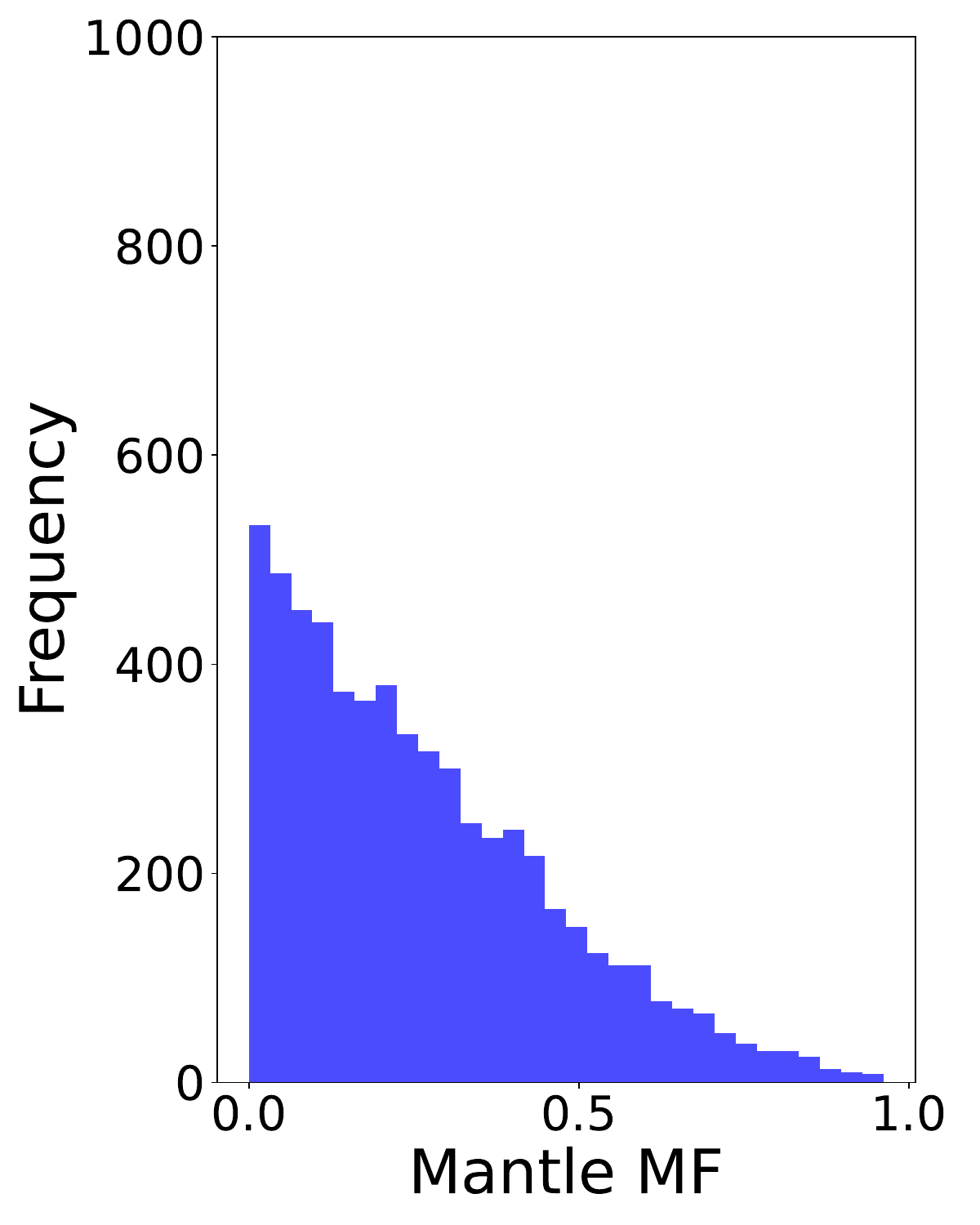}
        \includegraphics[width=0.25\linewidth]{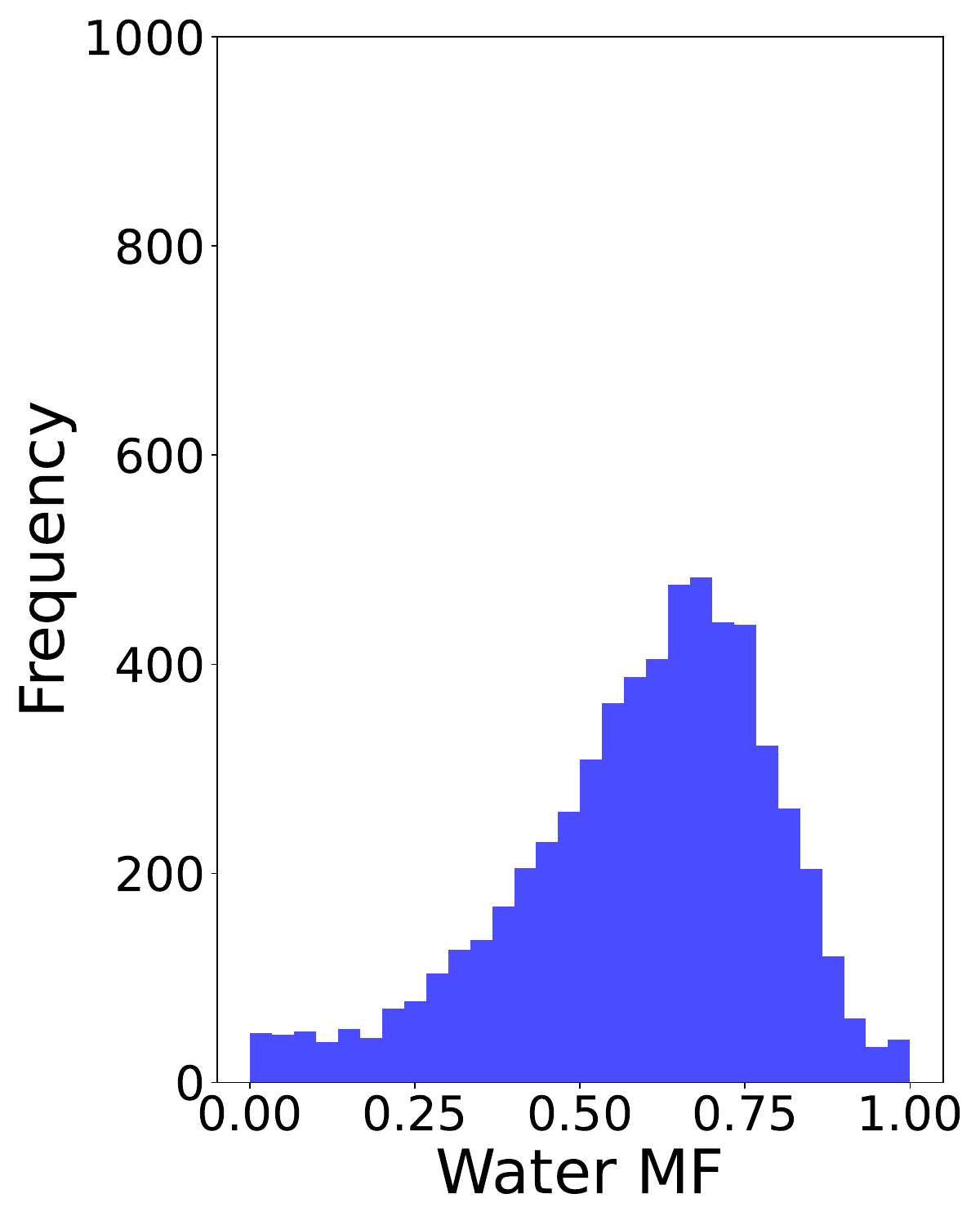}
        \includegraphics[width=0.25\linewidth]{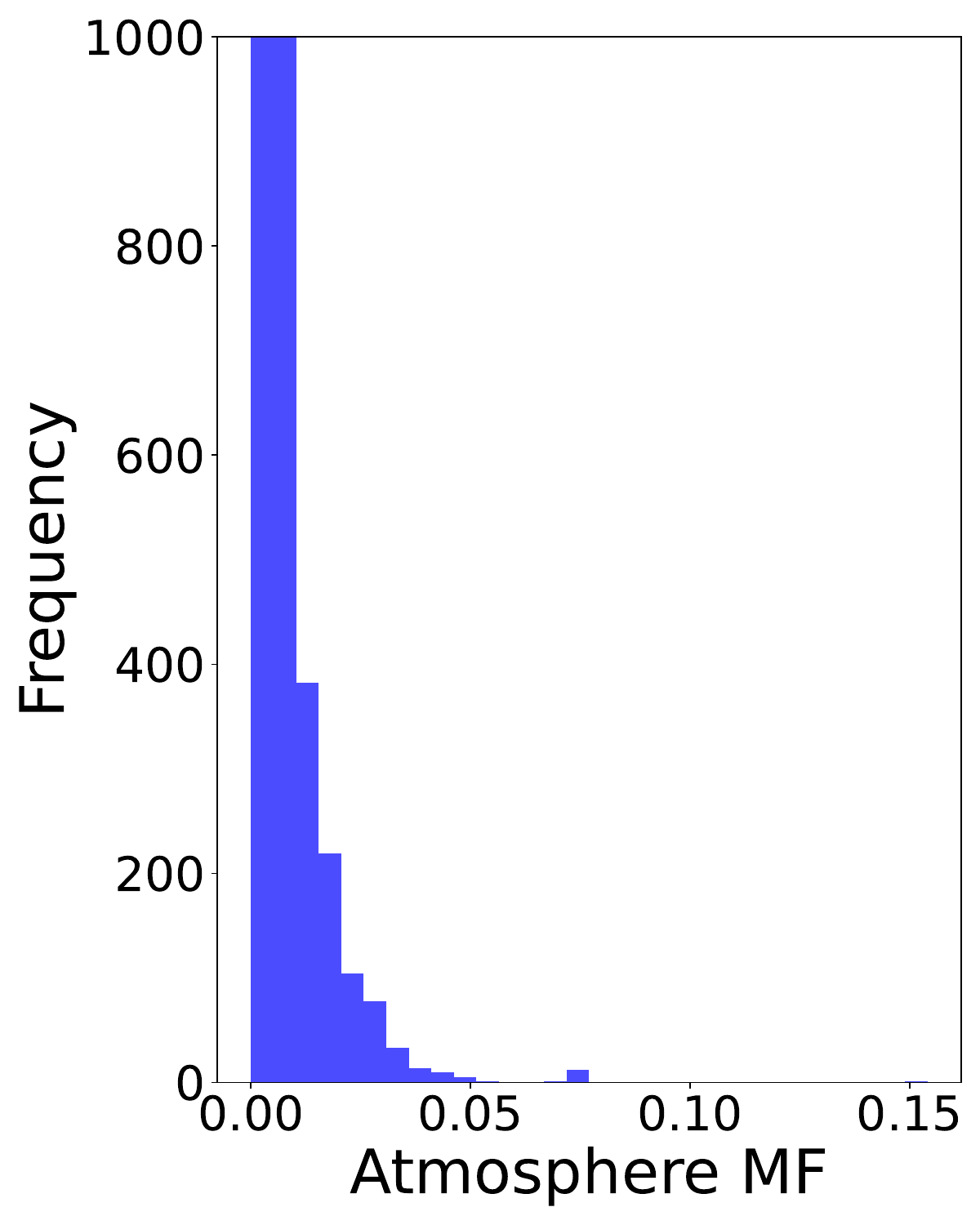}
        \subcaption*{LP 791-18 c}
    \end{minipage}

    \begin{minipage}{\linewidth}
        \includegraphics[width=0.25\linewidth]{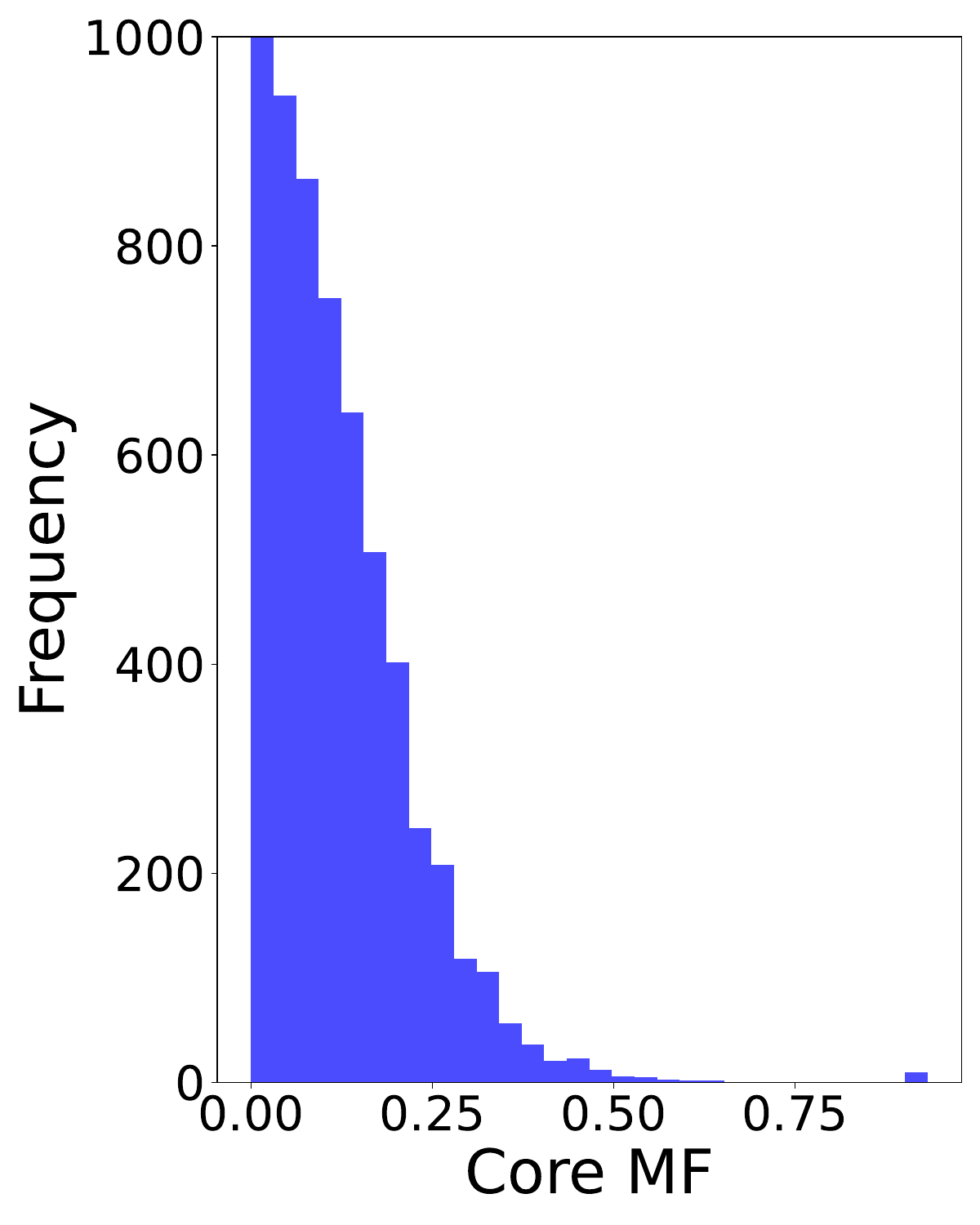}
        \includegraphics[width=0.25\linewidth]{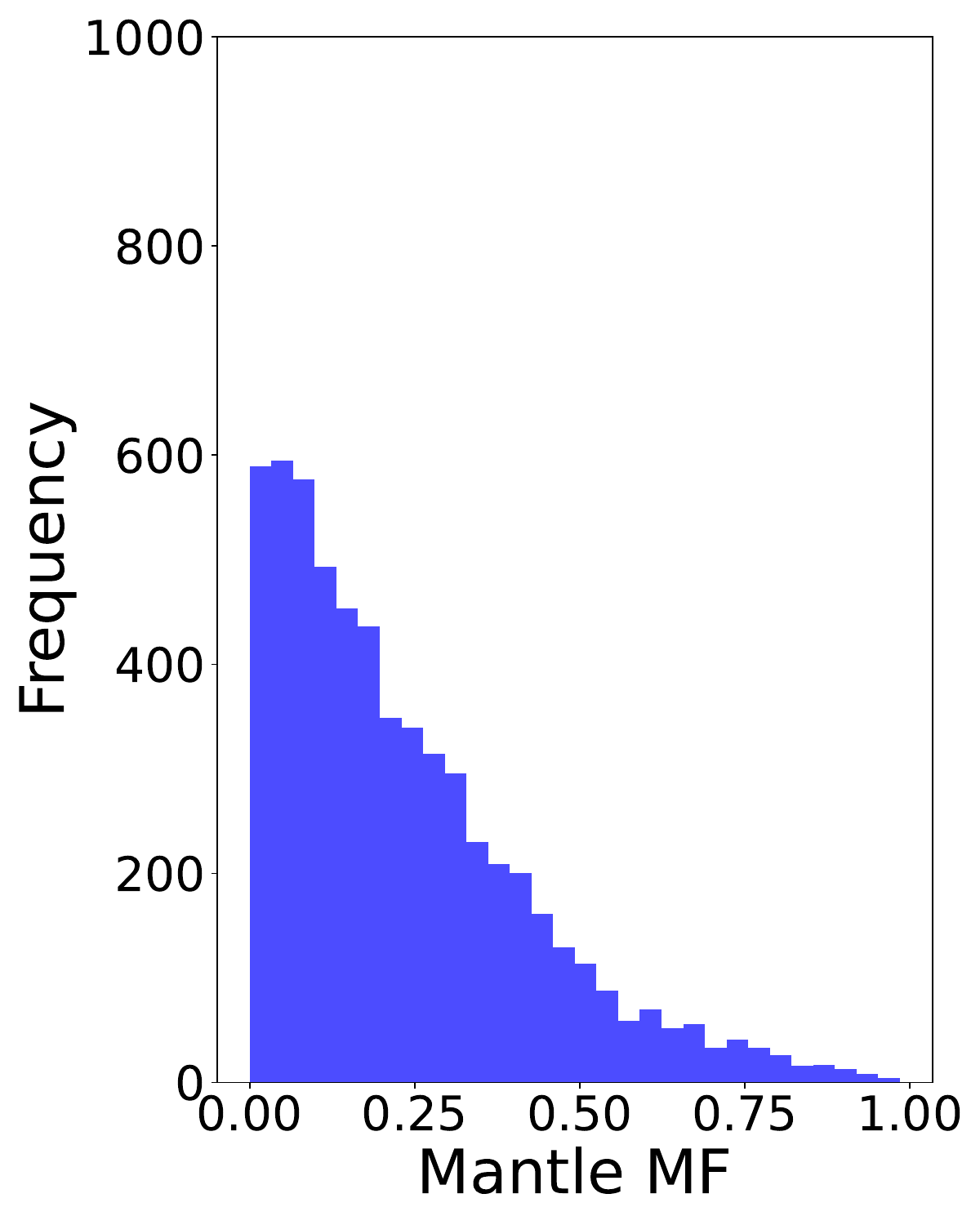}
        \includegraphics[width=0.25\linewidth]{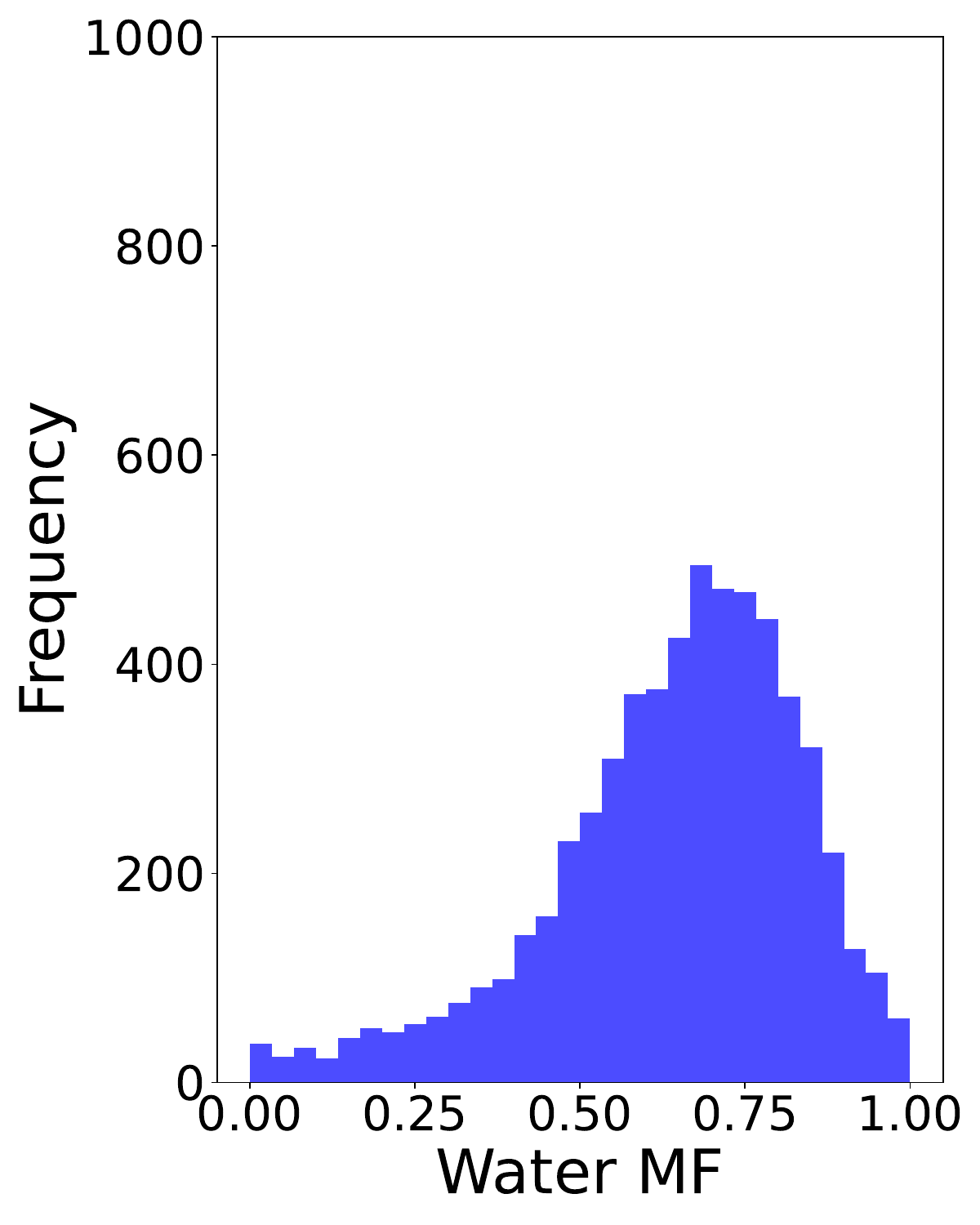}
        \includegraphics[width=0.25\linewidth]{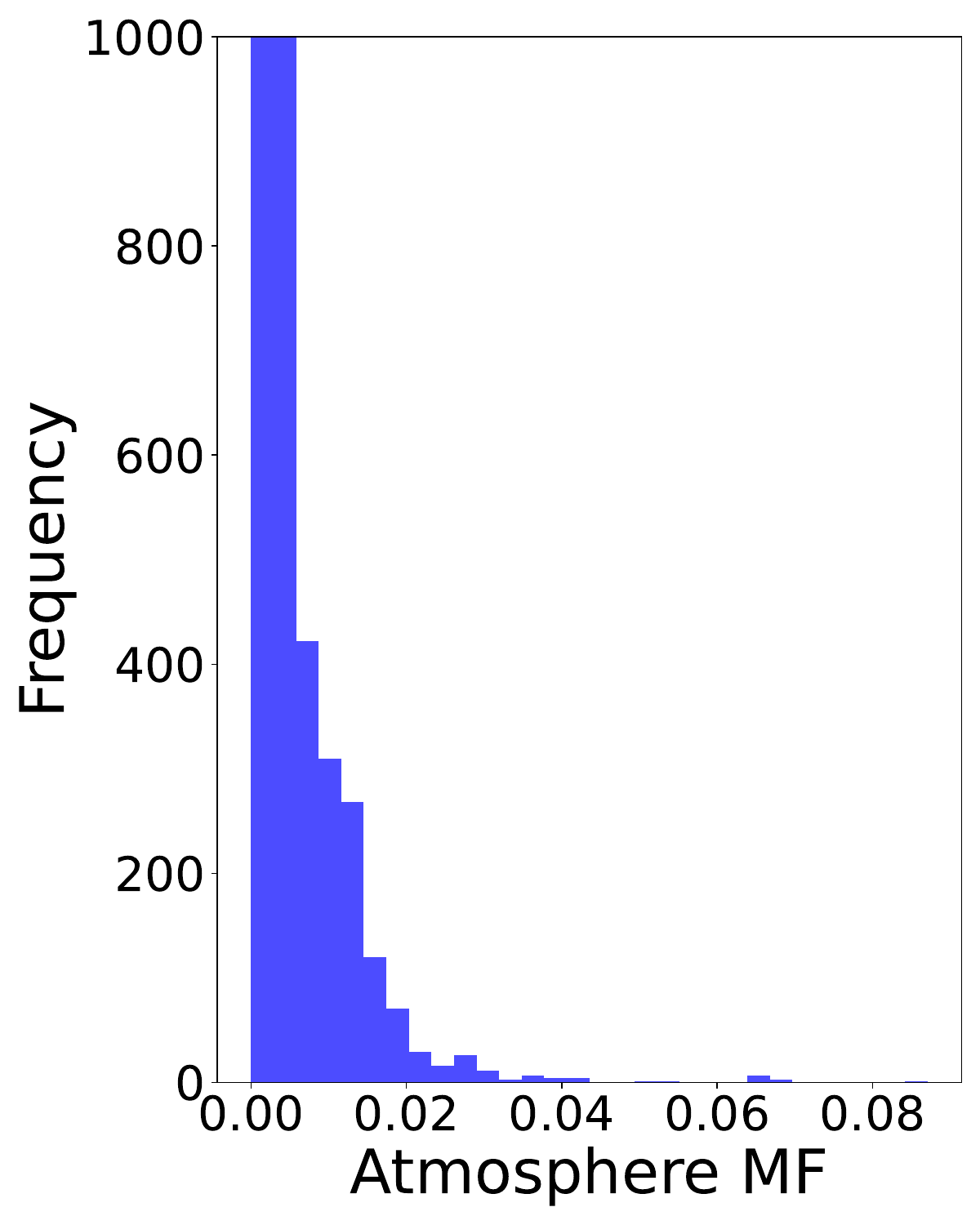}
        \subcaption*{LTT 3780 c}
    \end{minipage}

    \begin{minipage}{\linewidth}
        \includegraphics[width=0.25\linewidth]{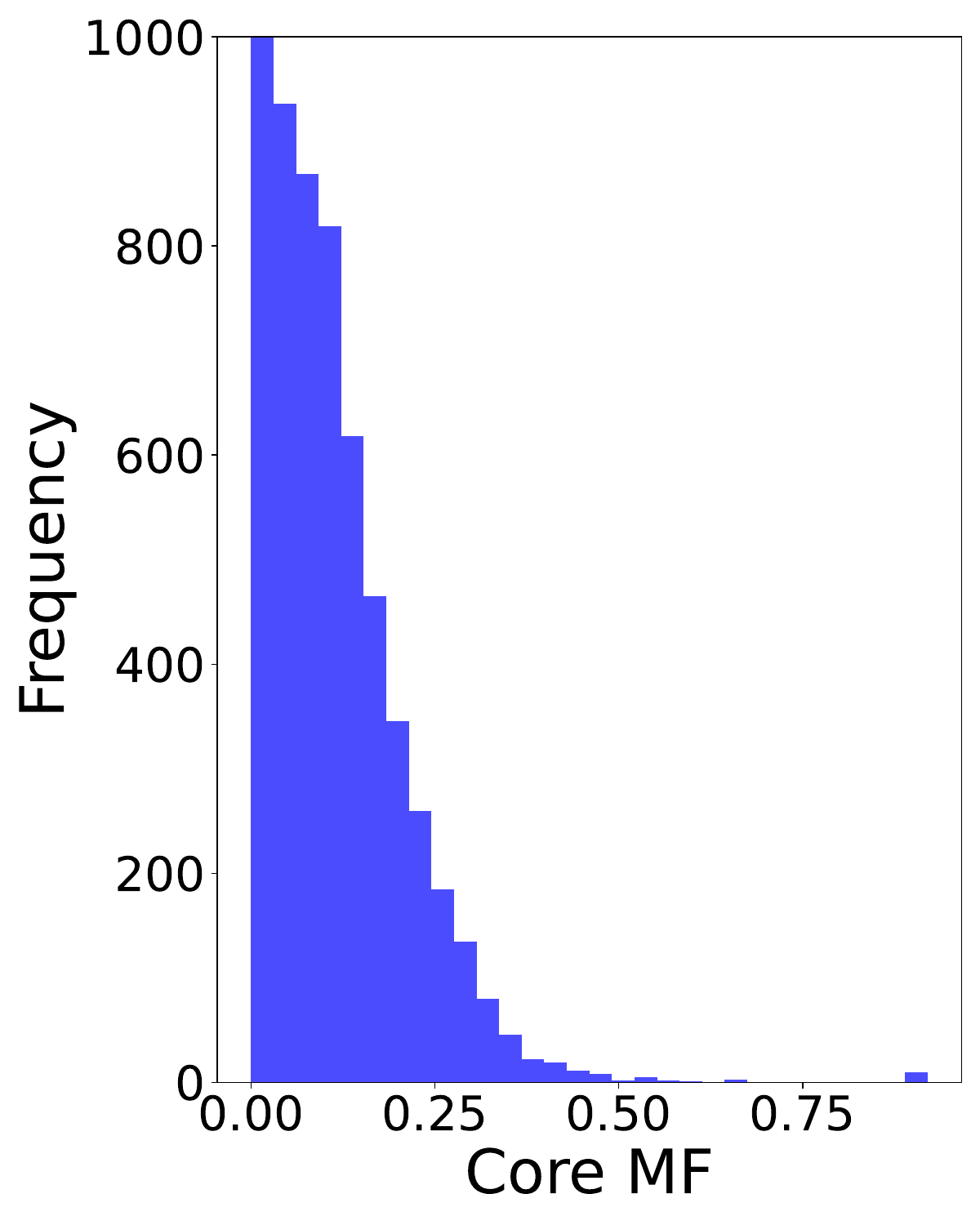}
        \includegraphics[width=0.25\linewidth]{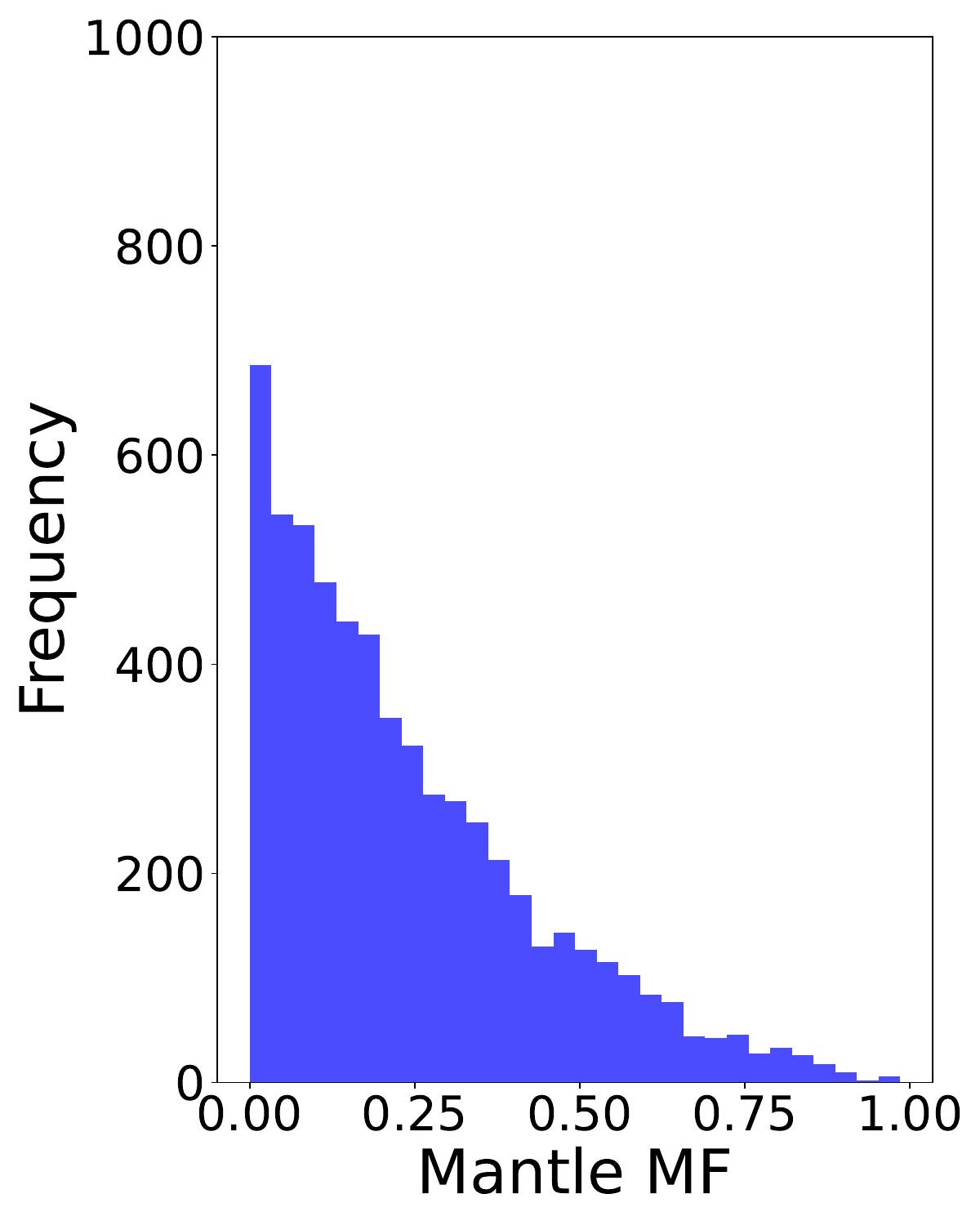}
        \includegraphics[width=0.25\linewidth]{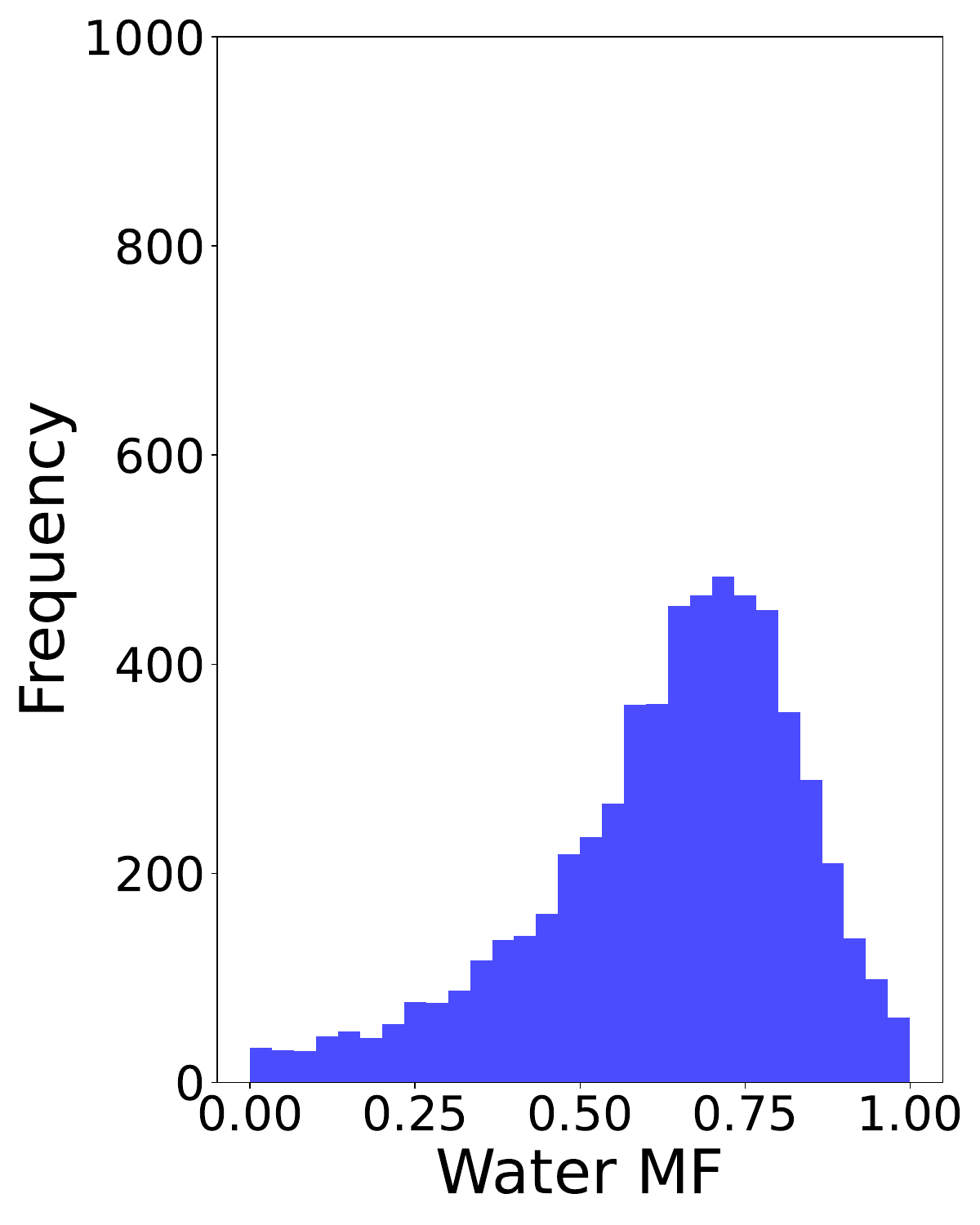}
        \includegraphics[width=0.25\linewidth]{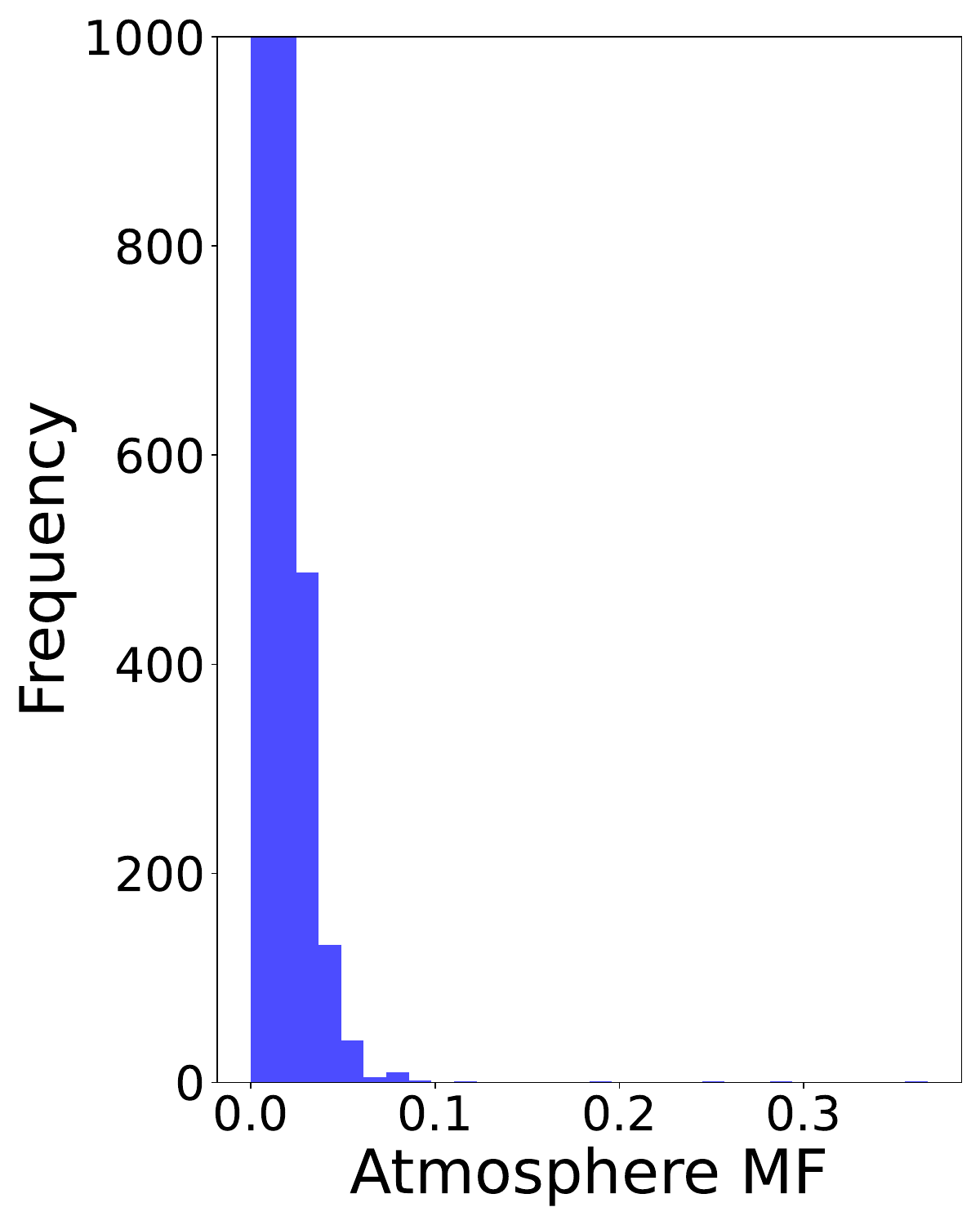}
        \subcaption*{K2-18 b}
    \end{minipage}

    \caption{(2/2) Continued from previous page.}
\end{figure*}

\section{Results}\label{3}
In this section we discuss the results obtained from the structural and atmospheric analysis of the individual exoplanets.

\subsection{Possible Structural Composition of the Exoplanets}\label{3.1}
\textbf{LHS 1140 b}: The interior structure analysis of the planet resulted in  $9^{+10}_{-9}$$\%$ of the core, $89^{+8}_{-11}$$\%$ of mantle, $0.5^{+0.4}_{-0.1}$$\%$ of water and 0$\%$ mass fraction of the atmosphere (see Figs.~\ref{fig:5}, \ref{fig:6}). However, $\sim$90$\%$ rocky composition of the exoplanet can not satisfactorily justify its observed density \citep{cadieux2024new} and also the locus of the planet on the mass-radius diagram (Fig.~\ref{fig:3}). Fig.~\ref{fig:4} highlights that a predominant rocky composition is not supported for this planet, and rather an Earth-like composition with $20-30\%$ of the core mass fraction can succumb to the observed bulk density of the planet. Similar conclusion can also be drawn from the histograms of the core and mantle mass fraction, CMF and MMF respectively, where we see tails towards higher CMF and lower MMF. Moreover, Fig.~\ref{fig:4} also supports the presence of small amounts of H-He-rich envelope above the core of this planet. Although, the planet has similar density to that of Earth, a precise $k_2$ measurement can also provide better results of interior analysis in agreement with all the observational constraints.

\textbf{TOI-1452 b}: Having a mass, radius, and bulk density similar to that of LHS 1140 b, this planet also shows a predominantly rocky composition ($89^{+12}_{-10}$$\%$) with only 9$\%$ of iron core and a negligible amount of water and atmosphere mass fraction. Taking into consideration the observed bulk density, this planet should also be composed of higher core mass fraction regardless of the presence of any gas envelope (see the locus of the planet on the radius-period space in Fig.~\ref{fig:4}). Tails towards higher CMF and lower MMF in both the histogram and the ternary plots also suggests an Earth-like composition for the planet. Alternatively, a small tail towards the higher WMF, points at the possibility of some amount of volatile being present in the planet.

\textbf{TOI-1266 c}: In agreement with its position in the mass-radius plot, TOI-1266 c shows a water rich interior with $79^{+46}_{-8}$$\%$ of water mass fraction (WMF), $4^{+22}_{-4}$$\%$ of CMF, $7^{+46}_{-8}$$\%$ of MMF, and negligible amount of atmosphere mass fraction (AMF). The large uncertainties in the mass fractions indicate that such high WMF cannot be a possible composition, and rather a lower WMF and higher MMF can better describe the interior structure of this planet. \citet{cloutier2024masses} also suggested for a water rich composition for the exoplanet. Considering its small observed mass corresponding to a large radius, an accurate fluid love number can precisely constrain the mass fractions of this planet. 

\textbf{LP 791-18 c}: similar to TOI-1266 c, this planet also shows a high fraction of the water composition with only $12^{+12}_{-08}$$\%$ of CMF, $23^{+25}_{-16}$$\%$ of MMF, and very small amount of AMF. Given it's observed mass and radius, this exoplanet should be rich with a small H-He gas layer above a 50-50 water and Earth-like core. The histogram tails towards lower WMF and higher MMF also speaks for a similar composition.  

\textbf{LTT 3780 c}: Consistent with the observed mass and radius, this planet also shows a high WMF of $67^{+23}_{-42}$$\%$, CMF of $9^{+21}_{-9}$$\%$, MMF of $18^{+49}_{-17}$$\%$ and $0.3^{+0.1}_{-2}$$\%$ AMF with a tail towards higher AMF. Both the histogram and the ternary plot supports a 50$\%$ water + 50$\%$ Earth-like composition with a small AMF enveloped around the core.

\textbf{K2-18 b}: This exoplanet shows a high AMF of $1^{+2}_{-1}$$\%$ with WMF of $66^{+23}_{-43}$$\%$, CMF of $9^{+19}_{-9}$$\%$ and MMF of $18^{+45}_{-17}$$\%$. The large histogram tails suggests for higher CMF and lower WMF, in agreement with the predictions of \citet{madhusudhan2020interior}. The modeled AMF also aligns well with the results of \citet{madhusudhan2020interior} favoring a water-rock composition for the planet, covered with small amount of H-He envelope.

\subsection{Present Day Envelope Mass Fraction}
\begin{figure*}        
	\includegraphics[width=0.45\linewidth]{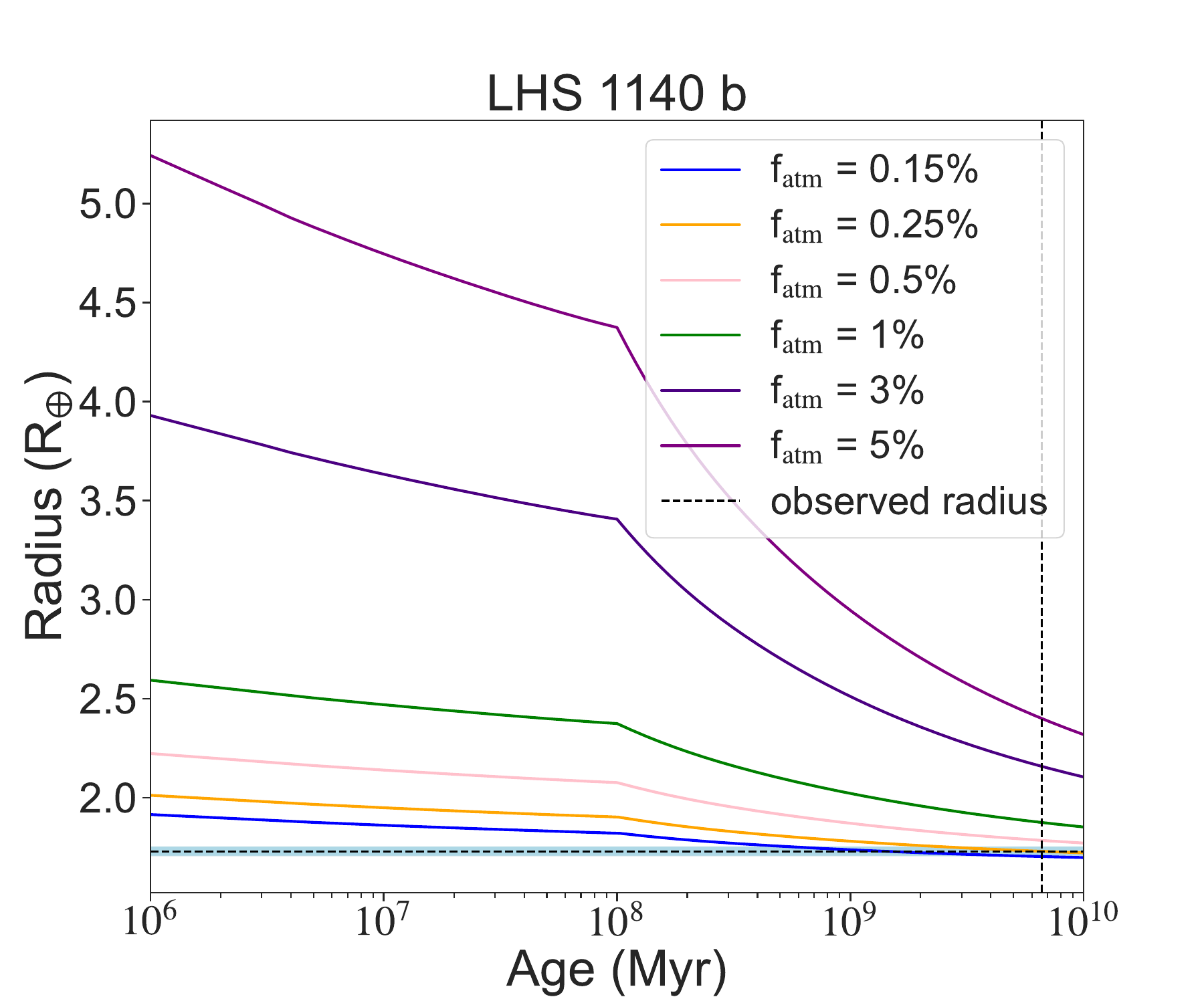}
    \includegraphics[width=0.45\linewidth]{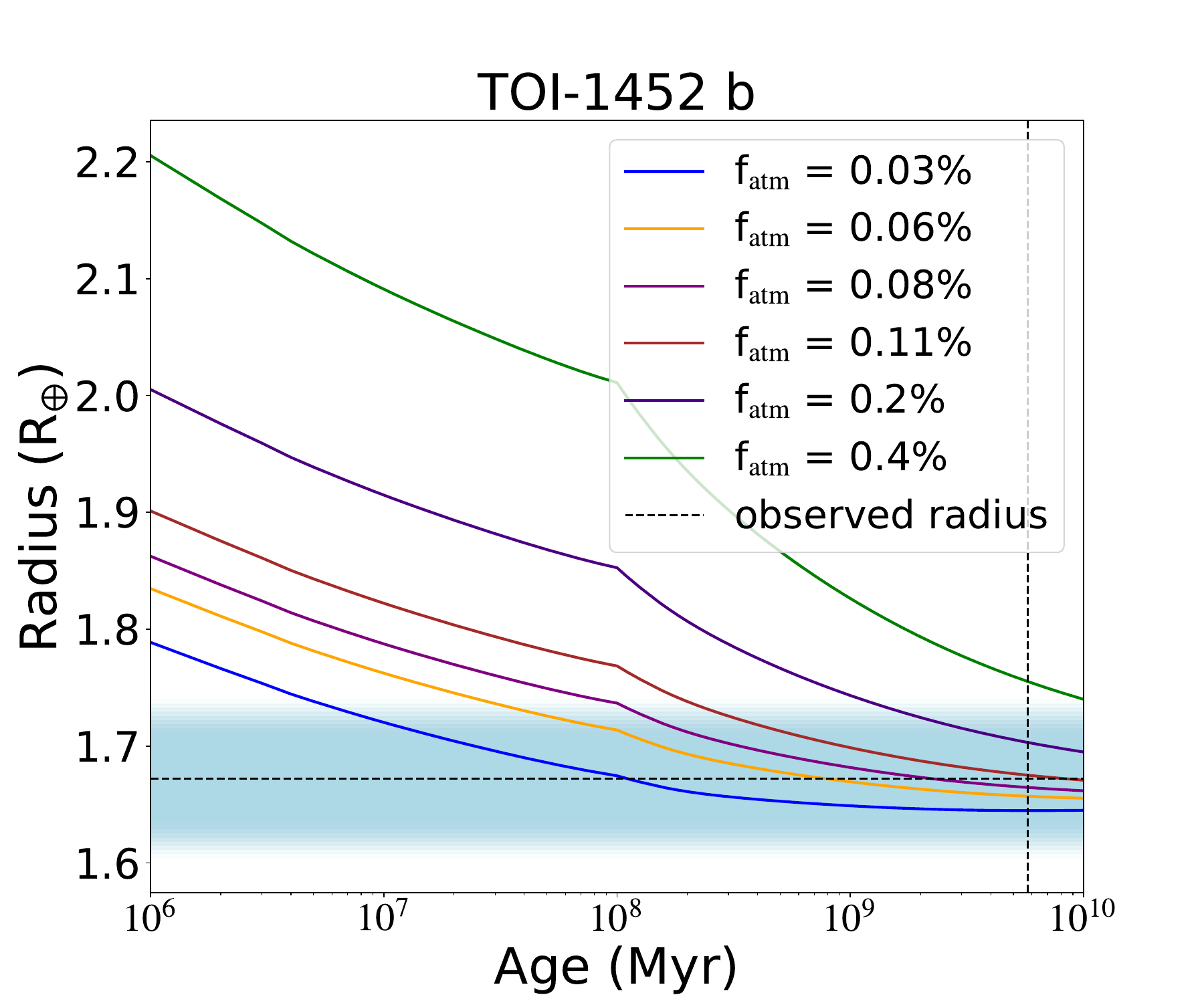}
	\includegraphics[width=0.45\linewidth]{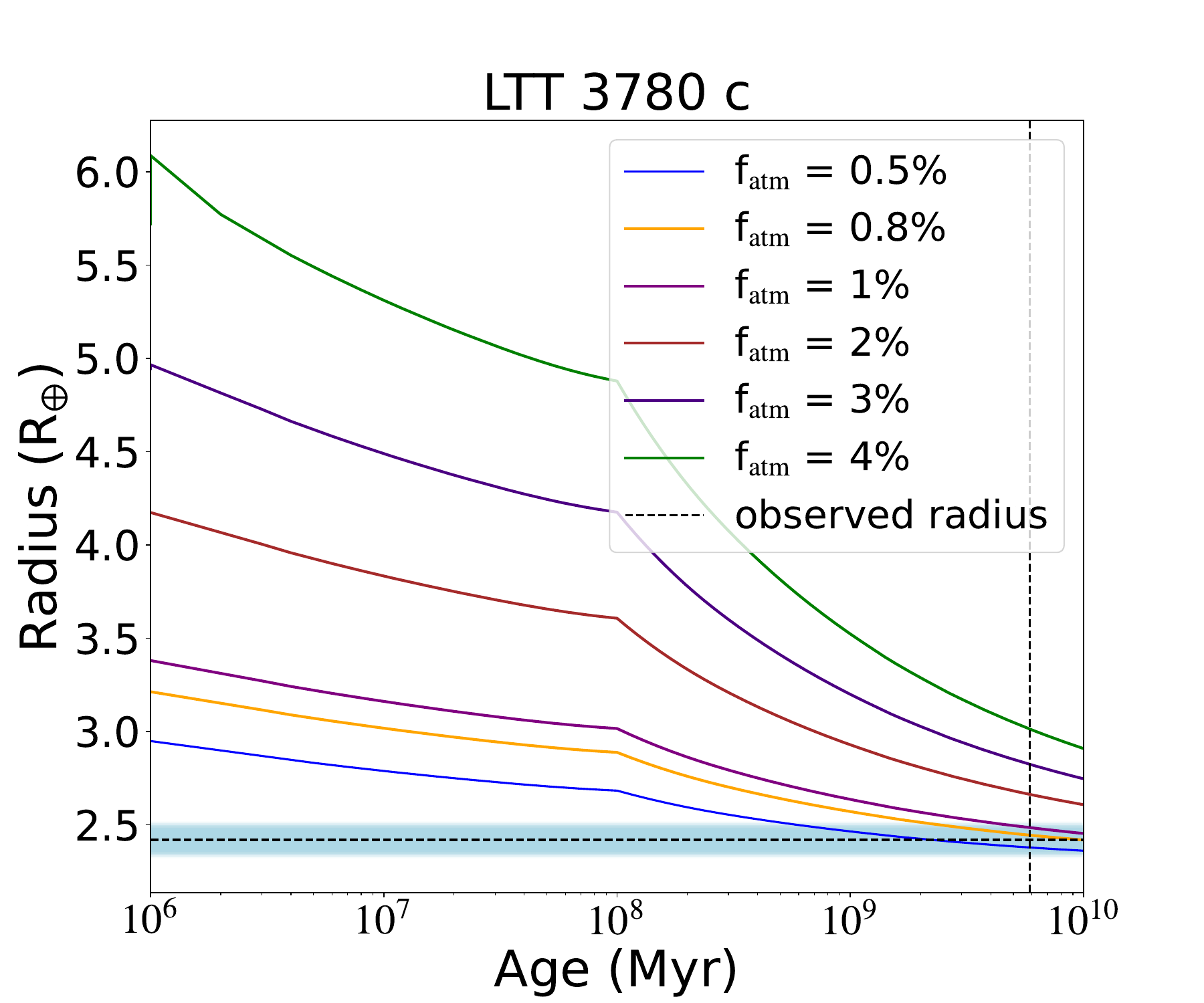}
	\includegraphics[width=0.45\linewidth]{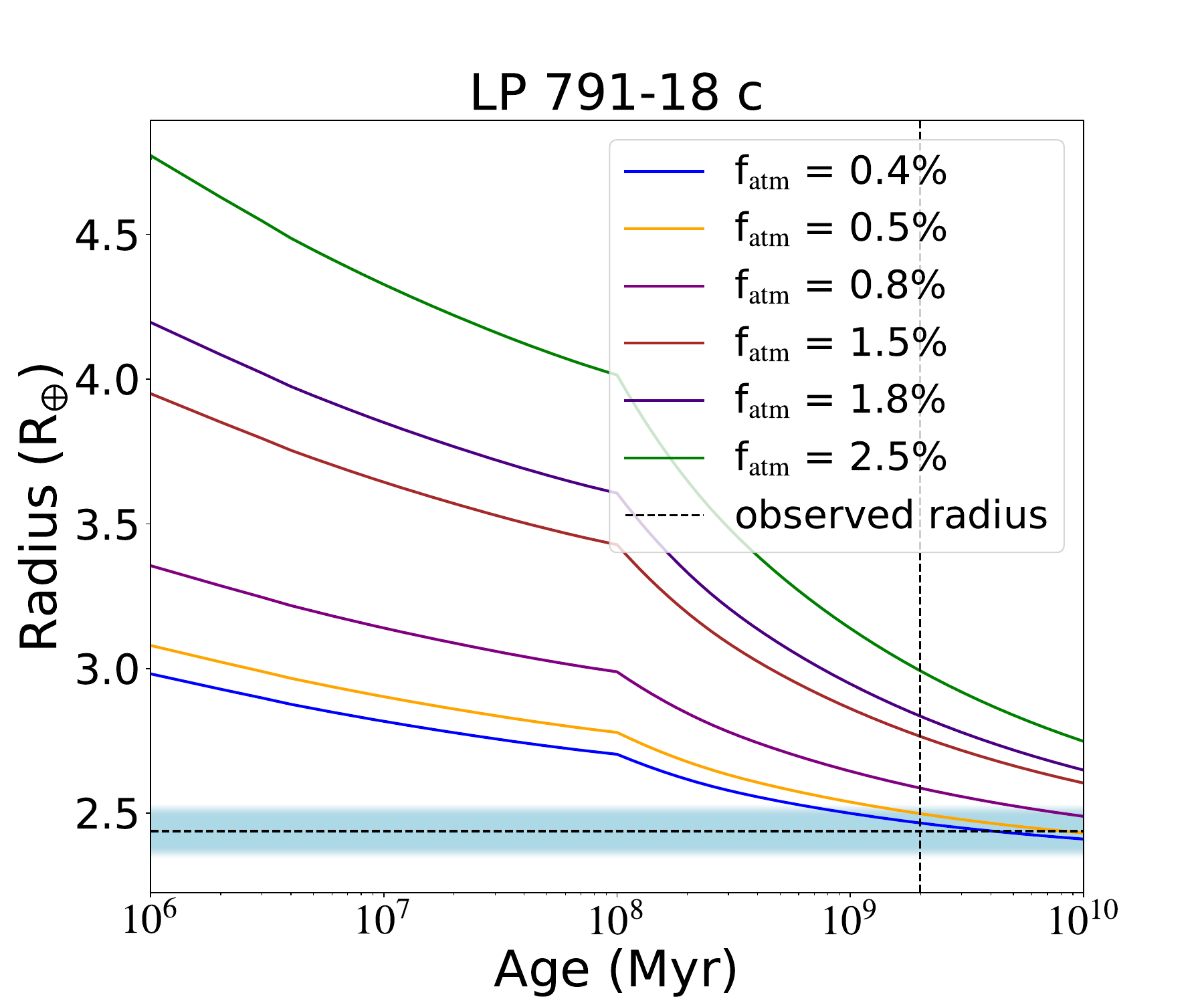}
	\includegraphics[width=0.45\linewidth]{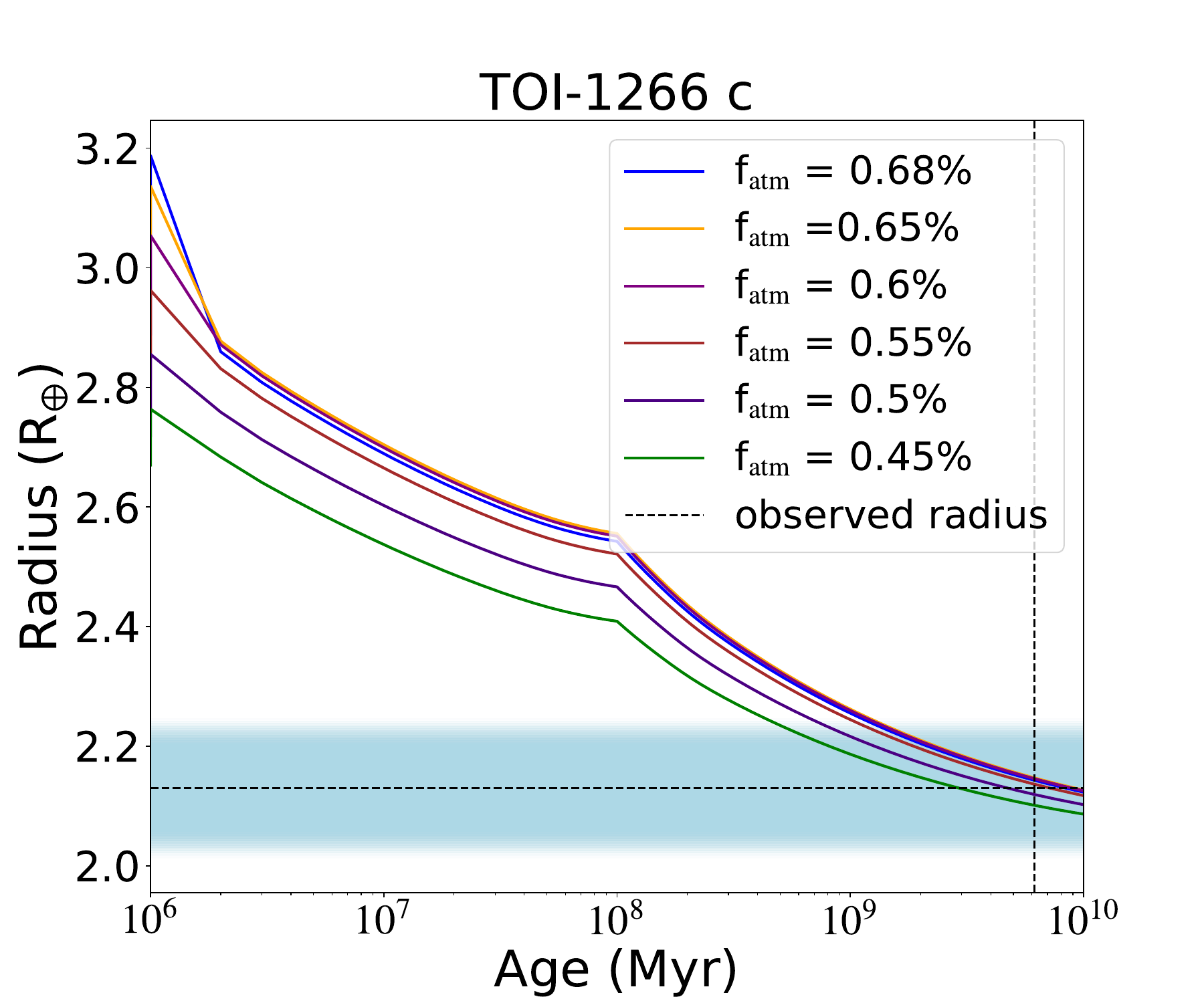}
	\includegraphics[width=0.45\linewidth]{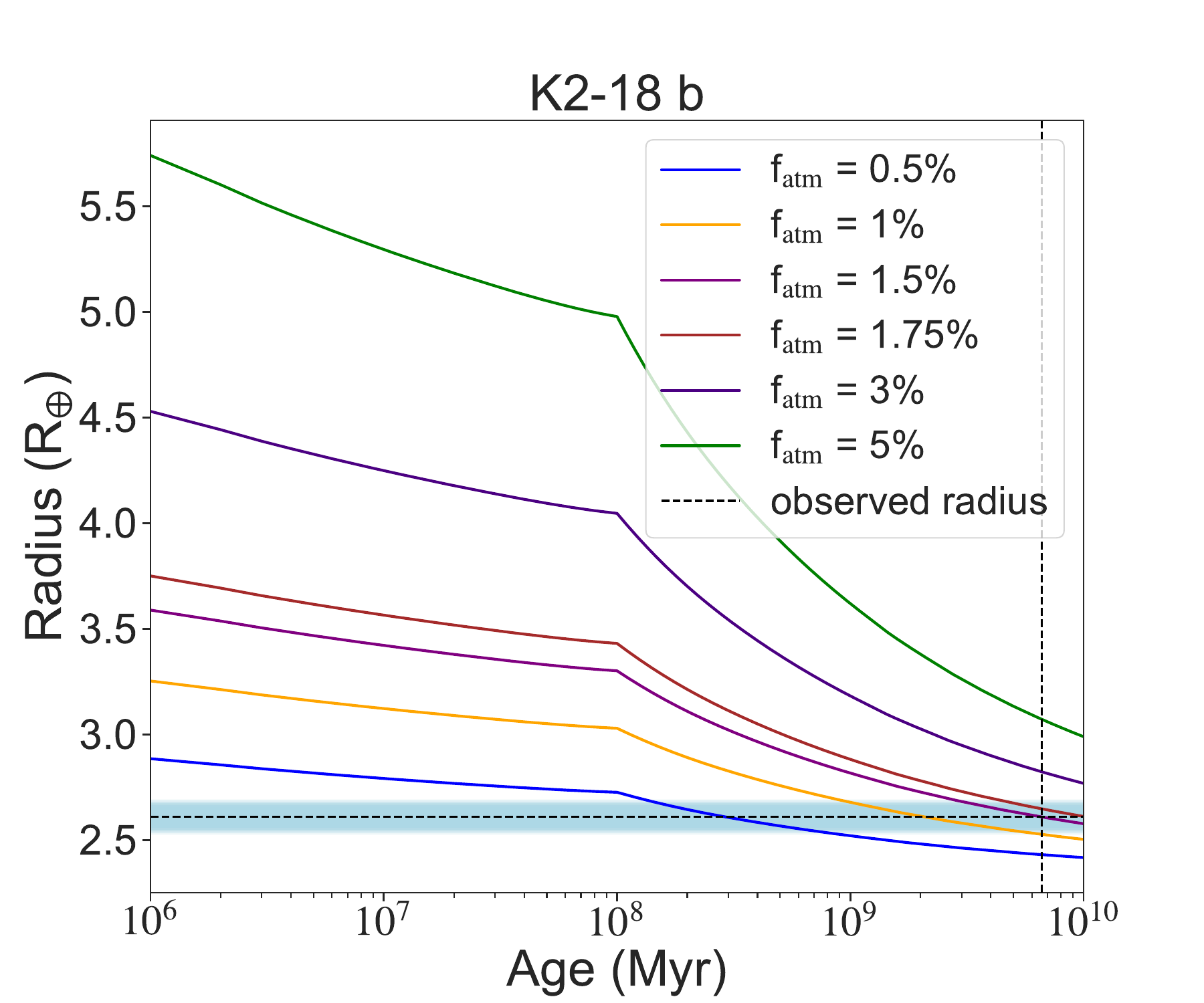}
    \caption{Photoevaporation simulation of the primordial H/He atmosphere for six exoplanets. The different curves represent various initial envelope mass fractions, as indicated in the legend. The vertical dashed line marks the estimated present age of the exoplanet system. The horizontal dashed line and the blue shaded region highlight the observed radius and its associated uncertainties, respectively.}
        \label{fig:7}    
\end{figure*}

As seen in Sec.~\ref{2.2} and Sec.~\ref{3.1}, LHS 1140 b and TOI-1452 b do not, observationally, satisfy a pure-rocky composition. Simulating the evolution of the atmospheric envelope of both exoplanets using rocky interior model of \citet{2019PNAS..116.9723Z}, also results in much smaller mass and radius for both the planets. We therefore, used the Earth-like (1/3 iron, 2/3 rock) composition for both the exoplanets and recovered their observed radius at different envelope mass fraction, provided the age of both the systems are more than 5Gyr (see Fig.~\ref{fig:7}). For LHS 1140 b, we find a present-day envelope mass fraction between $0.15\%$ - $0.25\%$ consistent with the observed mass and radius, closely reproducing the results from \citet{cadieux2024new}. The terrestrial composition of TOI-1452 b converges to a present day envelope mass fraction of around $0.1\%$ consistent with the prediction made by \citet{cadieux2022toi}. Both exoplanets show consistency with a mini-Neptune like composition with a very small amount of gaseous envelope atop their terrestrial cores. This small amount of atmosphere can also result from a gas-poor formation environment \citep{lee2021primordial}, where the planets were formed with very less amount of initial atmosphere. Although, from these simulation results we cannot confidently conclude on the formation mechanism of these planets, we can argue that a barren rocky build-up is also not a possible composition for both the exoplanets.

For rest of the water-rich planets, we used a 50$\%$ water+ 50$\%$ Earth-like composition model from \citet{2019PNAS..116.9723Z}. The mass loss mechanisms for all the planets are governed by photoevaporation and core-powered mass loss. From Fig.~\ref{fig:7}, we see that LTT 3780 c and LP 791-18 c are clearly consistent with the hydrodynamic escape model with their present day envelope mass fraction at $0.8\%$ and $0.4\,-\,0.5\%$, respectively. The dashed vertical line, denoting the present age of the planets, also closely supports the estimated mass fractions. These results are inconsistent with the `standard picture' of hydrodynamic escape model \citep{cloutier2024masses}, which rules out water or iron-rich core composition planets, around FGK stars, to be consistent with thermally driven mass loss mechanism \citep{gupta2019sculpting,owen2017evaporation}. The agreement of the water-rich composition of LTT 3780 c and LP 791-18 c with TDML mechanism argues that photoevaporation and core-powered mass loss can also be the formation and evolution mechanism for water-rich exoplanets around M-dwarf stars. The hydrodynamic escape model also shows that both these planets are possibly water-rich beneath a small atmospheric envelope. \\
For K2-18 b, we see that most of its atmosphere is stripped off and the model does not converge to a present-day envelope mass fraction consistent with the observed radius. Although the planet can sustain $1.75\%$ of its primordial envelope mass fraction if its age is more than or equal to 10Gyr (see Fig.~\ref{fig:7}). If the exoplanet is confirmed to be that old, it will probably have a water-rich core with an envelope mass fraction of $10^{-2}$, similar to the predictions of \citet{madhusudhan2020interior}. However, considering the planet to be younger than 10Gyr (see the vertical dashed line in the Fig.~\ref{fig:7}), the TDML models do not allow for the presence of atmospheric layer above the core of this planet. Alternatively, \citet{lozovsky2018threshold} argued that planets with radius similar to K2-18 b ($2.61R_{\oplus}$, \citet{benneke2019water}) should sustain a hydrogen rich atmosphere above a water-rich core and our interior structure analysis also supports the presence of small amount of atmosphere atop the core of the planet (see Fig.~\ref{fig:6}). Following the observations from our analysis and the arguments of \citet{lozovsky2018threshold}, K2-18 b will not be consistent with the TDML mechanism, suggesting a different formation and evolution pathway, which will allow the planet to sustain a H-He layer.

TOI-1266 c is clearly inconsistent with the TDML model as we could not retrieve the observed mass from the escape modeling. Both the present age of the planet, marked with the vertical line (see fig~\ref{fig:7}), and its interior analysis (see Figs.~\ref{fig:5} and ~\ref{fig:6}) do not support the presence of an atmospheric layer above the planet's core. \citet{cloutier2024masses} modeled the hydrodynamic escape of TOI-1266 c considering a Earth-like core and found the planet to be consistent with an envelope mass fraction of 3$\%$. They however, discarded this result and argued for a water-rich core composition of the planet to justify the demographics of the planetary system. From Fig.~\ref{fig:7}, we observe that this planet has completed stripping off its atmosphere and could barely reproduce the observed radius. Hence, only a water-rich core composition can justify the observed mass and radius of the planet.

The current envelope mass fraction, along with the derived mass and radius of the exoplanets, is presented in Table~\ref{tab3}. The radius uncertainties of the exoplanets (see Table~\ref{tab1}) are indicated by the blue-shaded region in all the plots in Fig.~\ref{fig:7}. 

\begin{table}
\centering
\caption{Present-day envelope mass fractions, along with the converged mass and radius values, for each individual exoplanet.}
\label{tab3}
\begin{tabular}{llll}
\hline
Planet & Envelope mass & Converged mass & Converged radius \\
   & $\%$  & ($M_{\oplus}$) & ($R_{\oplus}$)\\
\hline
LHS 1140 b   & $0.15 - 0.25$  & $5.673 - 5.678$  & $1.698 - 1.721$      \\
TOI-1452 b   & $0.11$         &$ 4.970 $          & $1.670$            \\
TOI-1266 c   & $0.60 - 0.68$   & $3.133$          & $2.120$             \\
LTT 3780 c   & $0.80$          & $6.237$          & $2.418$            \\
LP 791-18 c  & $0.40 - 0.50$    & $7.125$          & $2.431$            \\
K2-18 b      & $1.50 - 1.75$   & $8.623 - 8.640$     & $2.578 - 2.612$ \\ 
\hline
\end{tabular}
\end{table}

\section{Atmospheric characterisation prospect}\label{4}
The habitability of exoplanets is not solely determined by the presence of a water layer over a terrestrial core but also by the presence of an atmosphere, rich in atoms and molecules, enveloping the planet. Transit spectroscopy of the atmospheres of hot, inflated exoplanets using JWST and other space-based observations has revealed hydrogen- and nitrogen-rich compositions \citep[][and references therein]{constantinou2022characterizing,seidel2023detection,luque2020obliquity,damasceno2024atmospheric,macdonald2017hd}, while the atmospheres of cooler, temperate planets have been found to be rich in molecules such as H${_2}$O, CH${_4}$, NH${_3}$, CO, and CO${_2}$ \citep{madhusudhan2020interior,edwards2020hubble}.

Exoplanets around M-dwarfs exhibit deep transits due to their higher planet-to-star size ratio and serve as good targets for atmospheric characterisation by JWST and other space-based telescopes \citep{ridden2023high}. Exoplanets with equilibrium temperatures in the range $200 \leq T_\textrm{eq} \leq 400$ K that orbit relatively bright M-dwarfs are known to produce the highest transit signal strengths \citep{cadieux2022toi}. These atmospheric signal strengths can be characterized using the Transmission Spectroscopic Metric (TSM), as defined by Eq. 1 in \citet{kempton2018framework}. This metric depends on the planet's mass, radius, and equilibrium temperature, as well as the host star's radius and J-band magnitude. Exoplanets with low equilibrium temperature between $200 \leq T_\textrm{eq} \leq 400$ K, having high TSM values, are likely to have inflated radii and low masses, indicating the presence of extended, volatile-rich atmospheres. These planets, with strong signal strength, serve as best targets for transmission spectroscopy. Planets with radii $1.5<R_{p}<10 R_{\oplus}$ and $TSM \gtrsim 90$ have high SNR for atmospheric chracterisation and molecular detection with facilities such as JWST, ARIEL and ELTs \citep{kempton2018framework}. Table~\ref{tab4} provides a list of all the required parameters and the calculated TSM values for the sample exoplanets. We present precise TSM measurements for the exoplanets, based on updated mass and radius values. Among the studied exoplanets, LHS 1140 b and K2-18 b are two well-characterized temperate exoplanets, exhibiting signal strengths of 67.6 and 48.6, respectively. TOI-1452 b exhibits the lowest signal strength, though it is comparable to that of K2-18 b, making it a prime target for atmospheric characterization in upcoming JWST cycles.  TOI-1266 c has a transit strength (90.14) very close to the TSM cutoff for JWST follow-up. This also makes TOI-1266 c a favourable target for transit spectroscopy. The strongest signal strengths are observed for LTT 3780 c and LP 791-18 c, which makes them the most favorable targets, listed in the table, for future spectroscopic characterisation. 
\begin{table*}
  \centering
  \caption{Transmission spectroscopic metric (TSM) for the habitable zone temperate planets in our sample. Here $M_\textrm{p}$, \rp, and \teq{} represent the planetary mass, radius, and equilibrium temperature, respectively, while $R_{*}$ and $J$ denote the stellar radius and J-band magnitude of the host star. The sources of the parameters are listed in the final column.}
  \label{tab4}
\begin{tabular}{llllllllll}
\hline
Planet Name & $M_\textrm{p}$ & \rp{} & \teq{} & $R_{*}$  & $J$  & TSM & Source \\
    & ($M_{\oplus}$) & ($R_{\oplus}$) &(K)  & ($R_{\odot}$) & (mag)  &   &\\
\hline
LHS 1140 b   & 5.60     & 1.730    & 226       & 0.216     &   9.610       &  67.60  &   \cite{cadieux2024new}              \\
TOI-1452 b   & 4.82    & 1.672   & 326        & 0.275      & 10.604 &  39.90   & \cite{cadieux2022toi}             \\
TOI-1266 c   & 2.88    & 2.130    & 354         & 0.436      & 9.706        & 90.14   & \cite{cloutier2024masses}               \\
LP 791-18 c  & 7.10     & 2.438   & 370        & 0.171      & 11.559    & 158.70   & \cite{crossfield2019super,peterson2023temperate}             \\
LTT 3780 c   & 6.29    & 2.420    & 397         & 0.382       & 9.010      & 121.80   &\cite{nowak2020carmenes}              \\
K2-18 b      & 8.63    & 2.610    & 284          & 0.410       &  9.760         &              48.60  & \cite{2018AJ....155..257S,benneke2019water}       \\
\hline
\end{tabular}

\end{table*}

\section{Summary and conclusions}\label{5}
We conduct a detailed analysis of the habitability of 339 exoplanets, with radii and masses of $\leq 4R_{\oplus}$ and $\leq 15M_{\oplus}$, respectively. These constraints on mass and radius provide us with a subset of terrestrial extrasolar planets, ranging from super-Earths to mini-Neptunes. This subset of exoplanets enhances our focus on the habitability of smaller, terrestrial Earth-like planets, potentially sustaining some amount of primordial gaseous envelope. We plot these planets on the insolation flux versus stellar temperature diagram and identify 17 exoplanets located within the extended habitable zone (HZ) of their parent stars. All of these habitable exoplanets, except for Kepler-22 b, orbit M-dwarf stars and fall within the tidally locked radius of their host stars.

Along with the habitable planets, we also highlight the physical parameters of their host stars. The stars' spectral types are determined using Gaia photometry, along with various color-SpT and magnitude-SpT relations. We perform photometric SED fitting for each star and present the best-fit stellar parameters. We further assess the consistency of our study by thoroughly discussing and comparing our findings with existing archival literature.

Focusing on the habitable exoplanets, we examine their mass-radius relations and investigate their likely structure and composition. Among the 17 exoplanets, 12 have precise mass-radius measurements, revealing diverse possible compositions. The TRAPPIST-1 planets exhibit Earth-like terrestrial compositions, while the remaining planets show a transition from rocky super-Earth to mini-Neptune compositions. We also investigate the radius gap of exoplanets around M-dwarf stars, highlighting two distinct formation mechanisms responsible for the gap: TDML and GDF. We find two planets, LHS 1140 b and TOI-1452 b, situated within the radius valley, which supports different formation and evolutionary mechanisms. The formation mechanism of LHS 1140 b is debatable, as it does not align directly with any of the established formation curves (see Fig.~\ref{fig:4}). In contrast, TOI-1452 b lies on or slightly above the GDF curve, supporting a rocky, gas-depleted composition. However, further analysis of their interior structure does not fully support a purely rocky composition. Modeling their interior composition suggests a $20-30\%$ Earth-like composition. Photoevaporation simulations of their initial atmospheric envelopes also indicate small present-day envelope mass fractions.

The interior analysis of the remaining planets reveal water-rich compositions, accompanied by a small amount of envelope mass atop their cores. Notably, two water-rich core composition planets, LP 791-18 c and LTT 3780 c, show close alignment with the TDML mechanism, resulting in small envelope mass fractions consistent with their observed mass and radius. For K2-18 b and TOI-1266 c, we could not derive present-day envelope mass fractions consistent with their radii. Further observations and analyses are needed to confirm the presence of a gaseous layer atop the water or icy cores of these planets.

We finally investigate the atmospheric potential of the exoplanets and identify TOI-1266 c, with a signal strength near the JWST follow-up threshold, as a favorable target for future transit spectroscopy observations. Among all the planets, LP 791-18 c and LTT 3780 c exhibit the strongest transit signals, making them excellent targets for future observations.

\section*{Acknowledgements}
This research has made use of the NASA Exoplanet Archive, which is operated by the California Institute of Technology, under contract with the National Aeronautics and Space Administration under the Exoplanet Exploration Program. This publication also makes use of VOSA, developed under the Spanish Virtual Observatory (https://svo.cab.inta-csic.es) project funded by MCIN/AEI/10.13039/501100011033/ through grant PID2020-112949GB-I00. VOSA has been partially updated by using funding from the European Union's Horizon 2020 Research and Innovation Programme, under Grant Agreement nº 776403 (EXOPLANETS-A). A portion of the numerical computation of the work was performed on the computing cluster Pegasus of IUCAA, Pune, India. The authors express their sincere gratitude to Jorge Fern\'{a}ndez from the University of Warwick, UK, and Dr. Philipp Baumeister from Freie Universit\"{a}t, Berlin, for their generous help and guidance. SD extends here gratitude to Mangesh Daspute of Ariel University, for his suggestions and insights. Author SB highly acknowledges the Department of Science and Technology (DST), Govt. of India for providing DST INSPIRE fellowship vide grant no. IF220155. SD would like to thank her colleague Debojit Paul and Aalok Das for their unwavering support and motivation.

\section*{Data Availability}
All the necessary data required to generate the plots are included in the paper. The codes for modeling the interior and atmospheric evolution of the exoplanets are available in their respective GitHub repositories, with links provided in the footnotes wherever applicable.



\bibliographystyle{mnras}
\bibliography{example} 








\bsp	
\label{lastpage}
\end{document}